\documentclass[lettersize,journal]{IEEEtran}
\usepackage{amsmath,amsfonts}
\usepackage{algorithmic}
\usepackage{algorithm}
\usepackage{array}
\usepackage[caption=false,font=normalsize,labelfont=sf,textfont=sf]{subfig}
\usepackage{textcomp}
\usepackage{stfloats}

\usepackage{url}
\usepackage{breakurl}

\usepackage{verbatim}
\usepackage{graphicx}
\usepackage{cite}
\usepackage{amsthm}
\newtheorem{theorem}{Theorem}
\begin{document}

\title{\LARGE MStableChain: Towards Multi-Native Stablecoins in EVM-Compatible Blockchain for Stable Fee and Mass Adoption}

\author{
Mingzhe Li,~\IEEEmembership{Member,~IEEE,} 
Bo Gao,~\IEEEmembership{Member,~IEEE,} 
Kentaroh Toyoda,~\IEEEmembership{Member,~IEEE,} 
Yechao Yang, 
Juniarto Samsudin,
Haibin Zhang,
Sifei Lu,
Tai Hou Tng,
Kerching Choo, 
Andy Ting, 
Siow Mong Rick Goh,~\IEEEmembership{Senior Member,~IEEE,} 
and Qingsong Wei,~\IEEEmembership{Senior Member,~IEEE} 
\thanks{M. Li, B. Gao, K. Toyoda, Y. Yang, J. Samsudin, H. Zhang, S. Lu, T. Tng, S. Goh, and Q. Wei are with the Institute of High Performance Computing (IHPC), Agency for Science, Technology and Research (A*STAR), Singapore (email: li\_mingzhe@ihpc.a-star.edu.sg, gao\_bo@ihpc.a-star.edu.sg, kentaroh\_toyoda@ihpc.a-star.edu.sg, yang\_yechao@ihpc.a-star.edu.sg, juniarto\_samsudin@ihpc.a-star.edu.sg, zhang\_haibin@ihpc.a-star.edu.sg, lus@ihpc.a-star.edu.sg, tng\_tai\_hou@ihpc.a-star.edu.sg, gohsm@ihpc.a-star.edu.sg, wei\_qingsong@ihpc.a-star.edu.sg).}
\thanks{K. Choo, and A. Ting are with the Hela Labs, Singapore (email: kerching@helalabs.com, andy.ting@helalabs.com).}
}

\markboth{Journal of \LaTeX\ Class Files,~Vol.~14, No.~8, August~2021}%
{Shell \MakeLowercase{\textit{et al.}}: A Sample Article Using IEEEtran.cls for IEEE Journals}


\maketitle

\begin{abstract}
Traditional blockchain systems, such as Ethereum, typically rely on a \emph{single volatile cryptocurrency for transaction fees}. This leads to fluctuating transaction fee prices and limits the flexibility of users’ payment options. To address these issues, we propose MStableChain, which leverage multiple stablecoins as native tokens for transaction fee settlements, thus ensuring stable transaction fees and flexible payment options. To address the challenges of mass adoption and practicality, we propose several core designs.  To maintain compatibility with the Ethereum Virtual Machine (EVM) for mass adoption while supporting multiple native stablecoins, MStableChain employs a multi-currency units, multi-type RPCs mechanism. This mechanism enables the system to handle multiple stablecoins without altering the EVM or requiring changes to user applications. Furthermore, an oracle-based gas fee adjustment mechanism is proposed to manage exchange rates between different stablecoins, ensuring equitable transaction costs across various currencies. The system also introduces a secure, on-chain voting-based management protocol for the administrative functions related to these stablecoins. Experimental results from a prototype implementation demonstrate that MStableChain provides stable transaction fee prices, high effectiveness, and good usability.
\end{abstract}


\section{Introduction}

\IEEEPARstart{T}{he}
history of blockchain technology began with the creation of Bitcoin in 2008 by the anonymous figure Satoshi Nakamoto \cite{nakamoto2008bitcoin}. 
Bitcoin has opened up the broader use of cryptocurrencies. 
In 2015, Ethereum further revolutionized blockchain technology by introducing smart contracts, which are self-executing contracts with the terms directly written into code \cite{buterin2014next}.
After that, the emergence of numerous smart contract blockchains have expanded blockchain's applications beyond cryptocurrencies, enabling decentralized applications (dApps) and fostering innovation across various industries \cite{kalodner2018arbitrum, kanani2021matic, rocket2018snowflake}.
Many of these existing blockchain systems, especially public blockchains, issue their own \emph{\textbf{native}} cryptocurrency tokens \cite{nativetoken} (integrated at the system's base layer, such as ETH on Ethereum, which is distinct from smart contract tokens). 
Users of a particular blockchain system are required to \emph{pay transaction fees in its native token when sending transactions} \cite{kolb2020core}.

However, this model presents several significant issues. 
First, due to the volatility of the cryptocurrency market, the prices of issued tokens often fluctuate dramatically \cite{al2021cryptocurrency, saef2023regime, ndiaye2024blockchain, chen2023token}. 
This leads to substantial \emph{variability in the transaction fees} paid by users, which can severely affect the user experience \cite{zhang2024blockchain, Volatility2020}. 
For example, under similar network congestion conditions, Ethereum's transaction fees in 2021 could be more than ten times higher than in 2020 \cite{ethtxnfee}. 
Second, in the existing public blockchains, typically, only one native token is issued by a blockchain system for settling transaction fees \cite{wood2016polkadot, lokhava2019fast, buterin2014next, bnbchainwhitepaper, yakovenko2018solana, kalodner2018arbitrum}. 
Users must purchase the specific native token of a blockchain to perform subsequent transactions and pay associated fees. 
This requirement for users to buy a specific token is highly \emph{inflexible.}

To address the aforementioned research gaps, we propose MStableChain, a blockchain system that features relatively stable transaction fees and supports settlement in multiple native tokens. 
The core idea of MStableChain is to use \emph{various stablecoins as native tokens to settle transaction fees}, where the stablecoins are pegged to different fiat currencies (e.g., USD stablecoin, EUR stablecoin) \cite{mita2019stablecoin, zhang2023enhancing}.
First, by using stablecoins pegged to stable fiat currencies (instead of highly volatile cryptocurrency tokens) for transaction fee settlement, users can enjoy more \textbf{\emph{stable transaction fee prices}}. 
Second, the support for multiple native stablecoins enhances user \textbf{\emph{flexibility in payments}}, improving the overall user experience. 
For instance, European users can directly use EUR stablecoins to settle transaction fees without being forced to purchase and manage other tokens.

However, developing a practical and widely adopted blockchain system based on multiple native stablecoins is not straightforward. 
It faces several key challenges.

\vspace{6pt}
\noindent
\textbf{Challenge 1: \emph{How to support multiple native stablecoins for transaction fee settlement while balancing EVM compatibility for mass adoption.}}
EVM (Ethereum Virtual Machine) \cite{evm} compatibility is crucial for the widespread adoption of a blockchain system. 
One key fact is that among blockchains supporting smart contracts, those compatible with the EVM dominate the market, accounting for over 90\% of the Total Value Locked
according to data from DefiLlama \cite{chaintvl}.
However, EVM-compatible blockchains, whether from the perspective of the underlying system or upper-layer user applications, typically support only one native token for transaction fee settlement \cite{buterin2014next, kalodner2018arbitrum, kanani2021matic, bnbchainwhitepaper}. 
Modifying the EVM at the system level or adjusting user applications to support multiple native tokens would compromise EVM compatibility. 
Therefore, achieving both EVM compatibility and the support for multiple native token settlements presents a significant challenge.

To address this challenge, we design a \textbf{Multi-Currency Units, Multi-Type RPCs} mechanism to simultaneously satisfy the EVM compatibility for general users and support multiple native stablecoins. 
First, from the system's underlying perspective, to maintain EVM compatibility, we avoid making any changes to the EVM. 
Meanwhile, to support multiple native stablecoins, we add \emph{extra state fields for different currency units} into the account states that the blockchain system needs to maintain. 
These fields are used to record the state updates for different native stablecoins. 
Second, from the user application's perspective, to ensure EVM compatibility for ordinary users, we do not alter the applications. 
Simultaneously, to support multiple native stablecoins, we modify the communication channel between blockchain nodes (underlying system) and upper-layer applications: the RPC (Remote Procedure Call) \cite{li2021strong, luo2022last, kim2023etherdiffer, hara2020profiling}. 
We develop \emph{multiple types of RPC nodes, with each type corresponding to a different currency unit. }
A user can import the URL of the corresponding RPC into its wallet and use the native stablecoin corresponding to that RPC to settle transaction fees.

\vspace{6pt}
\noindent
\textbf{Challenge 2: \emph{How to manage exchange rates between different native stablecoins for transaction fee settlement.}}
With the introduction of multiple native stablecoins, different users might use different stablecoins to settle transaction fees. 
To ensure fairness, it is essential to manage the exchange rates between these stablecoins so that users spending different stablecoins incur equivalent costs. 
For example, if one user pays a transaction fee equivalent to 1 USD, another user might need to pay a fee equivalent to 0.93 EUR, depending on the current exchange rate. 
However, most existing systems support only single-token transaction fee settlements, making this issue particularly challenging to address.

To address this challenge, we propose a \textbf{Oracle-Based Gas Fee Adjustment} mechanism. 
The core idea is to use the oracle to obtain external exchange rate information and feed it to the underlying system, thereby adjusting the base fee to manage the exchange rates between different native stablecoins. 
First, to obtain external exchange rate information, we leverage the idea of the oracle. 
However, existing oracles \cite{pasdar2023connect, al2020trustworthy, gao2020cross, ezzat2022blockchain} typically provide data to smart contracts, whereas we need to feed information to the underlying system to control the exchange rates of native stablecoins. 
For the sake of universality, we choose to not alter the existing oracle services. 
Instead, we require the \emph{blockchain nodes to periodically synchronize the exchange rate information} from the oracle smart contract. 
In this way, the exchange rate information obtained by the oracle can be indirectly transmitted to the underlying system through the smart contract.
More importantly, to actually regulate the exchange rates, we use the \emph{exchange rate data from the oracle to adjust the base fee} (a core component of the transaction fee in EVM-compatible blockchains \cite{leonardos2021dynamical, liu2022empirical, roughgarden2021transaction, azouvi2023base}), thus altering the transaction fees users need to pay. 
In this way, when a user choose a specific native stablecoin to pay transaction fees, the system controls the transaction fees according to the exchange rate information provided by the oracle and adjusts the gas fee accordingly.

\vspace{6pt}
\noindent
\textbf{Challenge 3: \emph{How to securely and transparently manage  the on-chain minting and burning of native stablecoins.}}
Unlike most cryptocurrencies, which are issued according to rules set by the project team and can be arbitrarily created (e.g., ETH on Ethereum, BNB on Binance Chain), 
stablecoins require special attention due to their peg to fiat currencies, especially during on-chain minting and burning operations. 
Existing stablecoins are mostly smart contract tokens (e.g., USDT), which can be managed through smart contracts. 
However, since the stablecoins in our system are native tokens, they cannot be managed via smart contracts. 
Thus, the challenge lies in securely and transparently managing the on-chain operations of native stablecoins.

To address this challenge, we propose the \textbf{On-Chain Voting-Based Native Stablecoin Management} mechanism. 
We design a series of native stablecoin management functions including mint, burn, whitelist, etc. in the underlying system based on the idea of proposal and voting.
These functions are implemented as various types of native transactions (rather than smart contract transactions).
The entity backing a stablecoin can propose different types of management actions by submitting different proposal transactions (e.g., mint proposals) to the blockchain. 
Relevant participants (e.g., whitelisted individuals) can then cast their votes on these proposals through voting transactions. 
Since all management actions are recorded on-chain through transactions, this ensures transparency and the consensus mechanism guarantees security.

\begin{figure}[t]
\includegraphics[scale=0.57]{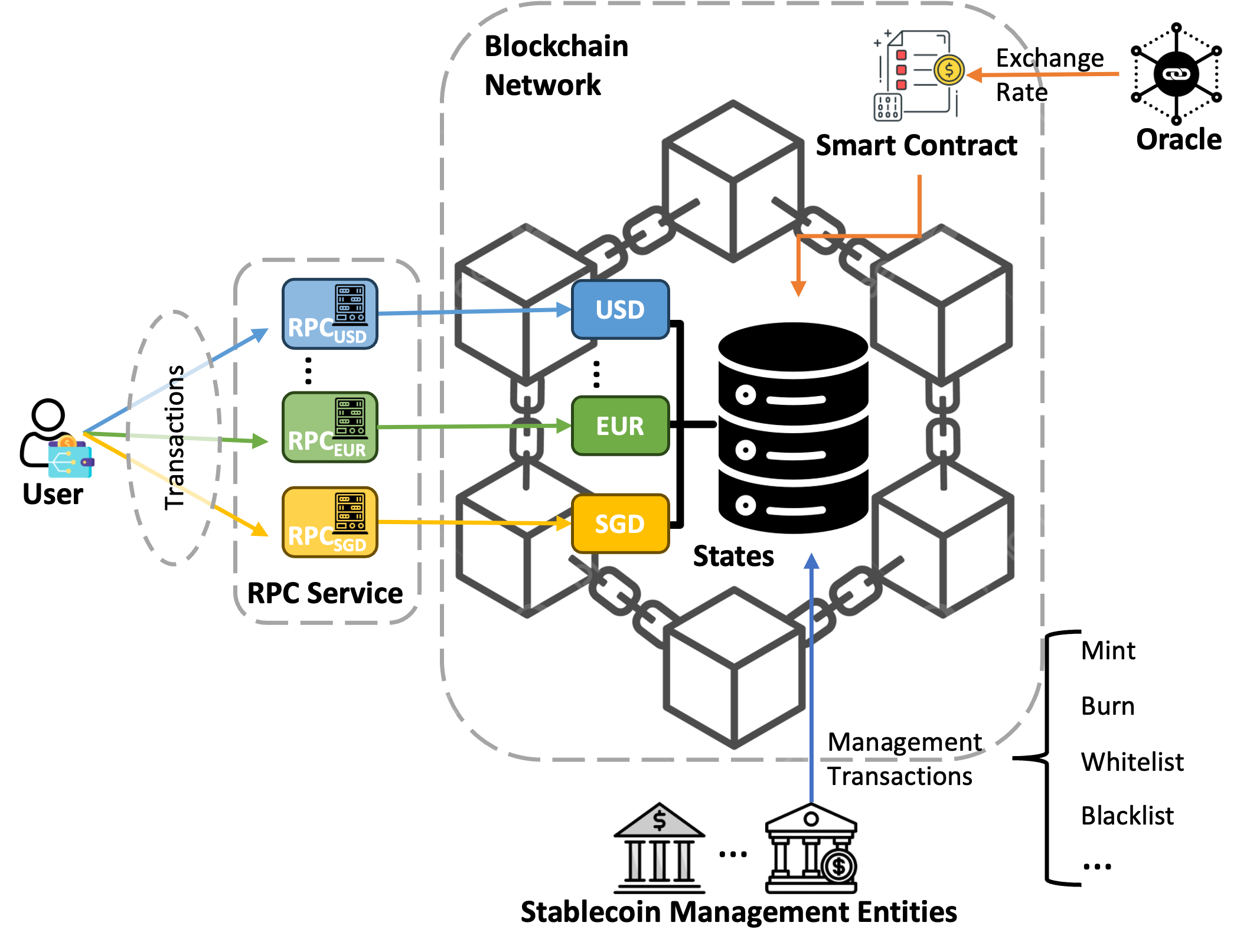}
\caption{Illustration of MStableChain's main idea. 
}
\label{fig:model}
\end{figure}

\vspace{6pt}
\noindent
\textbf{Contributions.}
With the resolution of the aforementioned significant challenges, MStableChain has been formally introduced. 
Figure \ref{fig:model} shows the general idea of MStableChain.
To the best of our knowledge, MStableChain is the first EVM-compatible blockchain that supports transaction fee settlements with multiple native stablecoins. 
It primarily \emph{addresses the issues of high transaction fee volatility and inflexible payment options} prevalent in many existing blockchain systems. 
Additionally, to ensure widespread adoption, MStableChain \emph{maintains EVM compatibility for general users. }
Within MStableChain, we design several key mechanisms to enhance the overall system. 
First, we achieve the support for multiple native stablecoins through a mechanism of Multi-Currency Units and Multi-Type RPCs without modifying the EVM or user applications. 
Second, the Oracle-Based Gas Fee Adjustment mechanism helps the system accurately manage exchange rates between different native stablecoins when settling transaction fees.
Third, the On-Chain Voting-Based Native Stablecoin Management mechanism enables secure and transparent management of native stablecoins by relevant entities.
We implement a prototype of MStableChain \cite{hela}, and the experimental results highlight the usability of MStableChain and show that it can provide stable transaction fee.

\section{Background and Related Works}

\subsection{Blockchain and Native Token}
\label{background_token}

Blockchain technology was first introduced with the creation of Bitcoin in 2008 by an anonymous figure known as Satoshi Nakamoto \cite{nakamoto2008bitcoin}. 
This innovation provided a decentralized, secure, and transparent way to conduct peer-to-peer transactions without the need for intermediaries. 
The technology further evolved with the introduction of Ethereum in 2015 by Vitalik Buterin \cite{buterin2014next}, which incorporated smart contracts, enabling the development of decentralized applications (dApps) and significantly expanding the potential uses of blockchain beyond digital currency transactions.

Most blockchain systems operate with their own native tokens, which are essential for the functioning of their respective networks \cite{kanani2021matic, kalodner2018arbitrum, rocket2018snowflake}. 
These native tokens are used to pay transaction fees, a fundamental aspect of maintaining and securing the blockchain \cite{kolb2020core}. 
For instance, Ethereum requires the use of Ether (ETH) to settle transaction fees \cite{buterin2014next}, while Binance Smart Chain uses Binance Coin (BNB) \cite{bnbchainwhitepaper}. 
This kind of system ensures that users must acquire and use the specific native token of a blockchain to perform transactions on that network.
However, this system model poses some problems, as will be described below.

\subsection{Volatile Transaction Fee and Inflexible Payment Options}
\label{background_volatile}

One significant issue with the aforementioned token mechanisms in existing blockchain systems is the volatility of native token prices, which leads to fluctuating transaction fees \cite{al2021cryptocurrency, katsiampa2019empirical, wang2023cryptocurrency, saef2023regime, ndiaye2024blockchain, chen2023token}. 
For example, the price of Ether (ETH) on the Ethereum network can vary significantly.
This has led to frequent fluctuations in transaction fees, with costs sometimes varying from several times to even tens of times more expensive in recent years \cite{ethtxnfee}.
This volatility makes it difficult for users to anticipate the costs of their transactions, potentially deterring them from using the network, especially during periods of high price fluctuation. 

Several efforts aimed at reducing transaction fee volatility have been proposed. 
However, most of them focus on the problem of transaction fee volatility due to network congestion or gas auction \cite{GSP2022, jin2023first, ferreira2021dynamic}. 
However, in this work, we mainly aim to address the problem of transaction fee price volatility due to token price fluctuations. 
Moreover, there are \emph{also mechanisms in MStableChain that can mitigate transaction fee fluctuations due to network congestion, as detailed in Section \ref{design_oracle}.}

Another problem with these systems is the inflexibility of payment options for transaction fees, as they only support a single native token \cite{wood2016polkadot, lokhava2019fast, buterin2014next, bnbchainwhitepaper, yakovenko2018solana, kalodner2018arbitrum}. 
This means that users must purchase and use the specific native token of the blockchain they wish to interact with, leaving them no other choice. 
For instance, to use the Ethereum network, a user must acquire Ether (ETH) to pay for transaction fees. 
This requirement can be inconvenient and cumbersome, as users are forced to buy and hold the specific token required by the blockchain, with no alternative payment options available. 
This lack of flexibility can be a barrier for users, especially those who prefer not to manage multiple cryptocurrencies. 

There have been a few attempts to address the issue of inflexible payment options for transaction fees, but each comes with its own limitations \cite{gsn, mtn}. 
A notable example is the Gas Station Network (GSN) \cite{gsn} proposed within the Ethereum community, which allows users to pay transaction fees using non-native tokens. 
GSN operates by having relayers forward users' transactions and temporarily cover the transaction fees with native tokens. 
The DApp developers then deduct the relevant tokens chosen by the user for transaction fee payment through a smart contract. 
However, the overall architecture of GSN is quite complex, requiring individual deployment by DApp developers. 
Moreover, the introduction of relayers increases the latency of each transaction, and the smart contract-based implementation increases the transaction fee cost to every transaction. 
This issue is also confirmed in our experimental section (Section \ref{subsec:gas_consumtion}). 

\vspace{6pt}
\noindent
\textbf{Motivation.}
As described above, the token mechanisms in existing blockchain systems face significant challenges, including transaction fee volatility and inflexible payment options. 
Moreover, current solutions have not effectively addressed these issues. 
This has motivated us to propose MStableChain, a blockchain system that supports transaction fee settlements with multiple native stablecoins. 
More importantly, MStableChain maintains EVM compatibility for ordinary users, facilitating widespread adoption. 
To ensure stable transaction fees, MStableChain uses stablecoins as native tokens for fee settlement. 
Additionally, MStableChain supports multiple native stablecoins at the system level, offering users flexible payment options without increasing transaction latency or cost.

\subsection{Stablecoin}
\label{background_stablecoin}

MStableChain uses native stablecoins to settle transaction fees.
Stablecoins are a type of cryptocurrency designed to maintain a stable value by being pegged to a reserve of assets, such as fiat currencies or commodities. 
Unlike traditional cryptocurrencies such as Bitcoin and Ethereum, which can experience significant price volatility, stablecoins aim to offer price stability.
Among stablecoins, fiat-collateralized stablecoins dominate the market (over 90\% of the market cap) \cite{stablecoinmarket}.  
These stablecoins are backed by fiat currency reserves held in a bank or other trusted institutions. 
For example, Tether (USDT) \cite{usdt} and USD Coin (USDC) \cite{usdc} are pegged to the US dollar, with each token representing one dollar held in reserve. 
This backing ensures that the stablecoin can be redeemed for the equivalent amount of fiat currency, maintaining its stable value. 

Unlike typical cryptocurrencies, stablecoins require additional management because they are pegged to fiat currencies. 
In existing blockchain systems, stablecoins are usually smart contract tokens (not native tokens) and are therefore managed by smart contracts.
These smart contracts automate the processes of minting new coins when fiat currency is deposited and burning coins when fiat currency is withdrawn. 
However, the stablecoins in MStableChain are native tokens at the system level, which cannot be easily managed through smart contracts. 
Therefore, we specifically design a series of native stablecoin management mechanisms at the system level to effectively govern these tokens.


\subsection{EVM Compatibility}
\label{background_EVM}

MStableChain accommodates both system-level and application-level EVM compatibility.
EVM (Ethereum Virtual Machine) compatibility \cite{evm, jia2022evm} refers to the ability of a blockchain system to execute smart contracts and run decentralized applications (dApps) in a manner consistent with the Ethereum network. 
The EVM is the runtime environment for smart contracts on Ethereum, enabling developers to deploy applications using a standardized set of rules and operations. 
This compatibility is crucial for ensuring that applications built on Ethereum can seamlessly operate on other EVM-compatible blockchains.

At the system level, EVM compatibility ensures that the blockchain can interpret and execute EVM bytecode, the low-level instructions processed by the Ethereum network. 
This compatibility facilitates the migration of projects and applications from Ethereum to other blockchains without requiring significant modifications to the underlying code.
To ensure EVM compatibility at the system level, MStableChain has chosen not to make any modifications to the EVM itself.

From the perspective of decentralized applications, EVM compatibility ensures that dApps can function correctly across different EVM-compatible blockchains. 
Developers can write smart contracts in Solidity, the primary programming language for Ethereum, and deploy them on any EVM-compatible network with minimal changes. 
Tools like MetaMask \cite{lee2019using}, which interact with the Ethereum network, can also seamlessly connect to other EVM-compatible blockchains, providing a consistent user experience across different platforms.
MStableChain also guarantees EVM compatibility at the application level. 
When ordinary users use MStableChain, their EVM-compatible applications (e.g. MetaMask wallets) do not need to be modified in any way.

\subsection{Other Preliminary Knowledge for Our Core Designs}
\label{background_other}

\noindent
\textbf{RPC.} Remote Procedure Call (RPC) is a protocol that enables communication between a client and a server across a network. 
In the context of blockchain, RPC is crucial for allowing users and applications to interact with the blockchain nodes \cite{li2021strong, luo2022last, kim2023etherdiffer, hara2020profiling}. 
Through RPC, clients can execute functions on the blockchain, such as querying the blockchain's state, sending transactions, and interacting with smart contracts. 
This interaction is facilitated by sending specific commands to the blockchain node, which processes these commands and returns the appropriate responses. 
In MStableChain, we design multiple types of RPCs, each corresponding to one of the native stablecoins in MStableChain. 
This helps common users to use EVM-compatible applications while still being able to interact with the multiple native stablecoins underlying MStableChain.

\vspace{6pt}
\noindent
\textbf{Oracle.} Blockchain oracles are services that provide external data to smart contracts on the blockchain, bridging the gap between off-chain data and on-chain applications \cite{pasdar2023connect, al2020trustworthy, gao2020cross, ezzat2022blockchain}. 
Since blockchains are inherently isolated systems, they cannot access data outside their network. 
Oracles solve this problem by fetching data from the real world—such as market prices, weather information, or sports results—and supplying it to the blockchain in a secure and reliable manner. 
This enables smart contracts to execute based on real-world events, expanding their functionality and potential use cases. 
In MStableChain, the oracle is used to obtain information about the exchange rate of each native stablecoin from an external source.

\vspace{6pt}
\noindent
\textbf{Gas Fee and Transaction Fee Components.} 
In EVM-compatible blockchains, the gas fees are typically calculated using the following formula (after Ethereum's EIP-1559 upgrade) \cite{leonardos2021dynamical, liu2022empirical, roughgarden2021transaction, azouvi2023base}: 

\begin{equation}
    Gas\ Fee = (Base\ Fee + Priority\ Fee) \cdot Gas\ Used.
    \label{equ:gas_fee}
\end{equation}

Gas fees are usually measured in the unit of Gwei or wei.
In the gas fee, the base fee is a mandatory component determined by the system, which typically adjusts according to the network congestion.
This fee is burned, effectively reducing the total supply of the native token and countering inflation. 
The tip (a.k.a., priority fee), on the other hand, is an optional additional fee that users can include to incentivize miners or validators to prioritize their transactions. 
The amount of gas used is related to the complexity of the transaction. 
Typically, a complex transaction will consume more gas.

On the other hand, how much worth of transaction fees a transaction ultimately consumes is also influenced by the price of a particular blockchain system's native token, as in the following formula:

\begin{equation}
    (Transaction\ Fee)_{USD} = Gas\ Fee \cdot (Token\ Price)_{USD}.
    \label{equ:txn_fee}
\end{equation}

As can be seen, even in the case of the same gas fee, the movement of the token price still causes fluctuations in the price of the final settlement transaction fee.
In MStableChain, to control the final transaction fee, it regulates the base fee based on the exchange rate information of the native stablecoin obtained by the oracle.

\section{System and Threat Model}

\subsection{System Model}
\label{model_system}

In MStableChain, there are $N$ blockchain nodes in the system, who are responsible for processing various types of transactions submitted by users, reaching consensus, and recording the transactions in the blockchain.
Like most practical blockchain systems \cite{buterin2014next, li2023lb, buchman2016tendermint, cason2021design, lokhava2019fast}, 
the blockchain nodes in MStableChain are connected by a partially synchronous peer-to-peer network, 
in which there exists an unknown global stabilization time after which all messages sent are delivered in less than a fixed amount of time \cite{li2023cochain}. 
As with many existing systems, each blockchain node has a unique public/secret key pair \cite{buchman2016tendermint, cason2021design}. 
A public key represents the identity of a node, which will be broadcasted through the network and recorded once a node joins.

MStableChain adopts the well-known Tendermint consensus protocol \cite{buchman2016tendermint}. 
However, the consensus protocol in MStableChain is actually modular and replaceable. 
The choice of consensus protocol does not fundamentally impact the core designs of MStableChain discussed in this paper. 
MStableChain can be applied in both public and consortium blockchain contexts (as discussed in Section \ref{analysis_generality}). 
In this paper, we apply MStableChain in a public blockchain scenario, using a PoS-based (Proof of Stake) version of the Tendermint consensus protocol. 
Without loss of generality, we assume that each blockchain node has an equal stake weight.
In addition, since MStableChain is EVM-compatible, our system also adopts the account model (similar to Ethereum) \cite{buterin2014next} to represent the blockchain state.
Each account has its own states and blockchain nodes record the states (e.g., balance) for all accounts. 

Blockchain systems typically include several external services, such as RPC and oracles \cite{crain2021red, li2023cochain, cason2021design, kiayias2017ouroboros, robinson2020layer, maram2021candid}, and MStableChain is no exception. 
The RPC service is generally used to handle requests from external applications calling the blockchain, and this is maintained by a group of RPC nodes. 
These RPC nodes can be provided by specialized third-party service providers or operated by the blockchain project itself. 
In MStableChain, the oracle is used to obtain exchange rate information from the external world and feed this data into the blockchain system. 
The oracle service is usually provided by dedicated oracle service providers \cite{breidenbach2021chainlink}.


\subsection{Threat Model}
\label{model_threat}

There are two kinds of blockchain nodes in MStableChain: honest and malicious. 
The honest nodes obey all the protocols. 
However, malicious
(Byzantine) nodes may corrupt the protocols in arbitrary manners,
such as arbitrarily packing invalid transactions into blocks (e.g., transaction
manipulation), sending messages with different values to different
nodes (e.g., equivocation attack), or failing to send any or all messages
(e.g., silence attack).
The fraction of total malicious blockchain nodes in the system is denoted as $F$, meaning $FN$ nodes are controlled by Byzantine adversaries in the whole system. 
To guarantee the secure operation of the blockchain system, the fraction of malicious blockchain nodes in the network $F$ needs to be less than the fault tolerance threshold of the consensus protocol (e.g., 1/3 for a BFT-typed consensus protocol, like Tendermint \cite{buchman2016tendermint}).

Since RPC nodes and the oracle are considered external services, existing works typically assume that they operate honestly \cite{crain2021red, li2023cochain, cason2021design, kiayias2017ouroboros, robinson2020layer, maram2021candid}. 
MStableChain adopts the same assumption. 
In fact, for the oracle, there are now many decentralized oracle services, such as Chainlink \cite{breidenbach2021chainlink}, that can resist the influence of malicious oracle nodes in a Byzantine environment and provide accurate data feeds. 
Regarding the RPC service, it is common practice to deploy multiple RPC nodes to enhance the resilience of the service. 
This redundancy ensures that the RPC service can continue to function properly even in the presence of faulty RPC nodes under normal circumstances.

\section{System Design}

\subsection{Overview}
\label{design_overview}

MStableChain is an EVM-compatible blockchain system that supports transaction fee settlements with multiple native stablecoins. 
It aims to address the issues of high transaction fee volatility 
and inflexible payment options prevalent in many existing blockchain systems. 
To achieve these goals, we propose three core designs. 
First, in Section \ref{design_multi}, we detail how the design of multiple currency units and various RPC types simultaneously ensures EVM compatibility for a broad range of users and supports transaction fee settlements with multiple native stablecoins. 
Second, in Section \ref{design_oracle}, we describe the proposed oracle-based gas fee adjustment mechanism, which is used to regulate the exchange rates between different native stablecoins during transaction fee settlements. 
Third, in Section \ref{design_mgmt}, we elaborate on a series of secure and transparent native stablecoin management mechanisms based on on-chain voting. 
In addition, in Section \ref{design_workflow}, we provide examples of the system's interaction processes to help readers better understand our system design.

\subsection{Multi-Currency Units, Multi-Type RPCs}
\label{design_multi}

\begin{figure}[t]
\hspace{-12pt}
\includegraphics[scale=0.54]{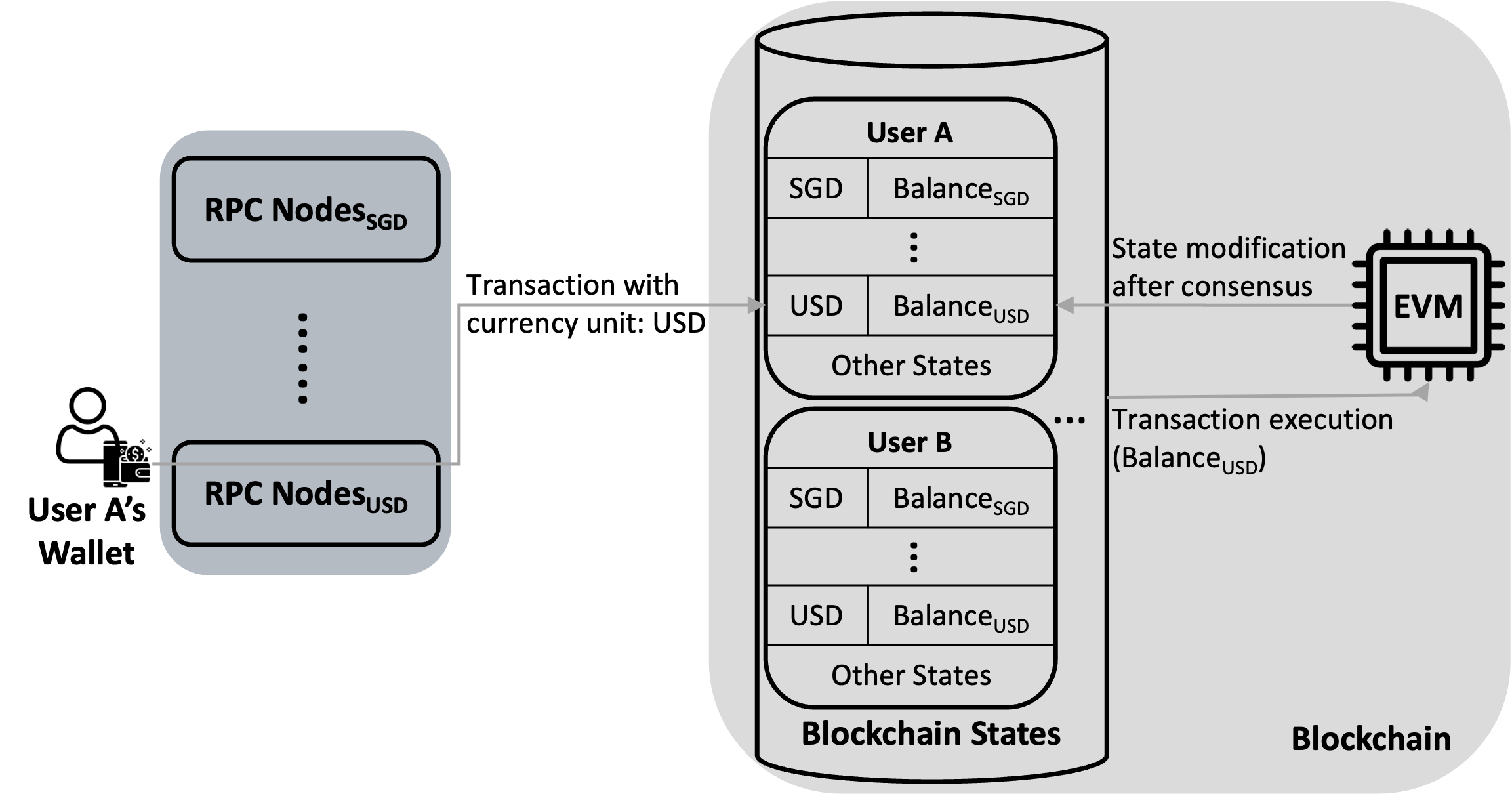}
\caption{Multi-Currency Units, Multi-Type RPCs. Take user A as an example.
}
\label{fig:multi}
\end{figure}

In this part, we focus on one of our core designs: Multi-Currency Units, Multi-Type RPCs. 
This design enables MStableChain to support transaction fee settlements with multiple native stablecoins, providing relatively stable transaction fee prices and flexible payment options. 
At the same time, it ensures EVM compatibility for a wide range of general users, facilitating widespread adoption of MStableChain.
The main idea of the design is illustrated in Figure \ref{fig:multi}.

\vspace{6pt}
\noindent
\textbf{Multiple Currency Units.}
Modifications to the underlying blockchain system are necessary to support multiple native stablecoin transaction fees while maintaining EVM compatibility. 
Our intuition is to achieve this by adding multiple fields to record currency units and their corresponding balances in the account state without altering the EVM.

Specifically, for other general EVM-compatible blockchains, the primary component of the account state stored by blockchain nodes is the account balance. 
Since EVM-compatible blockchains usually have only one native token, each account has a single balance to store its unique native token (note that this refers to the native token, not smart contract tokens stored by smart contract accounts). 
However, MStableChain needs to support multiple native tokens for transaction fee settlements. 
To achieve this, we add multiple balances to the account state stored by blockchain nodes and correspondingly add multiple currency units to differentiate between various stablecoins. 
Consequently, at the system level, each account has multiple balances for different stablecoins, distinguishable by different currency units.

For EVM-compatible blockchains, the logic of transactions is generally executed through the EVM. 
However, the EVM itself does not have the capability to distinguish multiple token balances. 
As a result, the challenge is how to allow EVM to handle multiple types of native stablecoins without modifying EVM itself.
To address this, we modify the underlying system so that when a node processes a transaction, it only passes the balance of the selected native stablecoin and other necessary information (e.g., account nonce) to the EVM for execution. 
The information regarding multiple currency units and balances of other non-selected native stablecoins is not passed to the EVM, ensuring EVM compatibility. 
After the transaction is executed by the EVM, the blockchain node updates the balance state of the corresponding native stablecoin.

\vspace{3pt}
\noindent
\textit{Remarks.} 
In the aforementioned design, a malicious blockchain node might intentionally confuse the balance updates of different native stablecoins when updating the account state. 
However, as will be discussed in Section \ref{design_oracle}, through our designs and the protection provided by the consensus, such malicious behavior can be prevented. 
This also guarantees the security of the system.

\vspace{6pt}
\noindent
\textbf{Multiple-Type RPCs.}
In the aforementioned design, we have achieved compatibility with both EVM and multiple native stablecoins at the system level. 
However, for general users, it is also important to \emph{ensure EVM compatibility at the application level} while supporting multiple native stablecoins to settle transaction fees.
Common EVM-compatible applications (e.g., various types of EVM-compatible wallets such as MetaMask, OKX Wallet \cite{lee2019using, okx}) cannot distinguish between multiple native tokens on the same blockchain, since EVM-compatible blockchains typically have only one native token. 
To enable applications to differentiate and use multiple native stablecoins in MStableChain without modifying the upper-layer applications (to maintain EVM compatibility), we propose the design of multiple types of RPCs, as detailed below.

In MStableChain, we design multiple types of RPCs to correspond to various currency units within the system.
We add one distinct currency unit to each type of RPC, with each type of RPC managing a specific currency unit. 
There are multiple sets of RPC nodes, with each type of RPC deployed on a set of RPC nodes operating separately. 
Additionally, each type of RPC has a unique set of URLs for user applications to connect to. 
In the user application (e.g., MetaMask wallet), users can select the type of stablecoin they wish to use by importing the URL of the corresponding type of RPC. 
Users can also import multiple types of RPC URLs and switch between them to use different types of stablecoins. 
When a user initiates a transaction using a specific stablecoin, the transaction is sent to the corresponding type of RPC. 
This type of RPC appends the corresponding currency unit information to the transaction and forwards it to the blockchain node. 
Once the blockchain nodes receive the transaction, they will process it according to the appended currency unit information.

\vspace{3pt}
\noindent
\textit{Remarks.} 
It is important to note that the above design does not require any changes to the user application, but rather makes reasonable use of some of the existing functionalities in the application.
In current wallets, users can change the RPC URL they connect to, typically used to switch between different blockchain networks. 
In MStableChain, we use this feature to switch between different types of RPC nodes, allowing users to use different types of native stablecoins in the same blockchain.

\subsection{Oracle-Based Gas Fee Adjustment}
\label{design_oracle}

\begin{figure*}[t]
\centering
\includegraphics[scale=0.63]{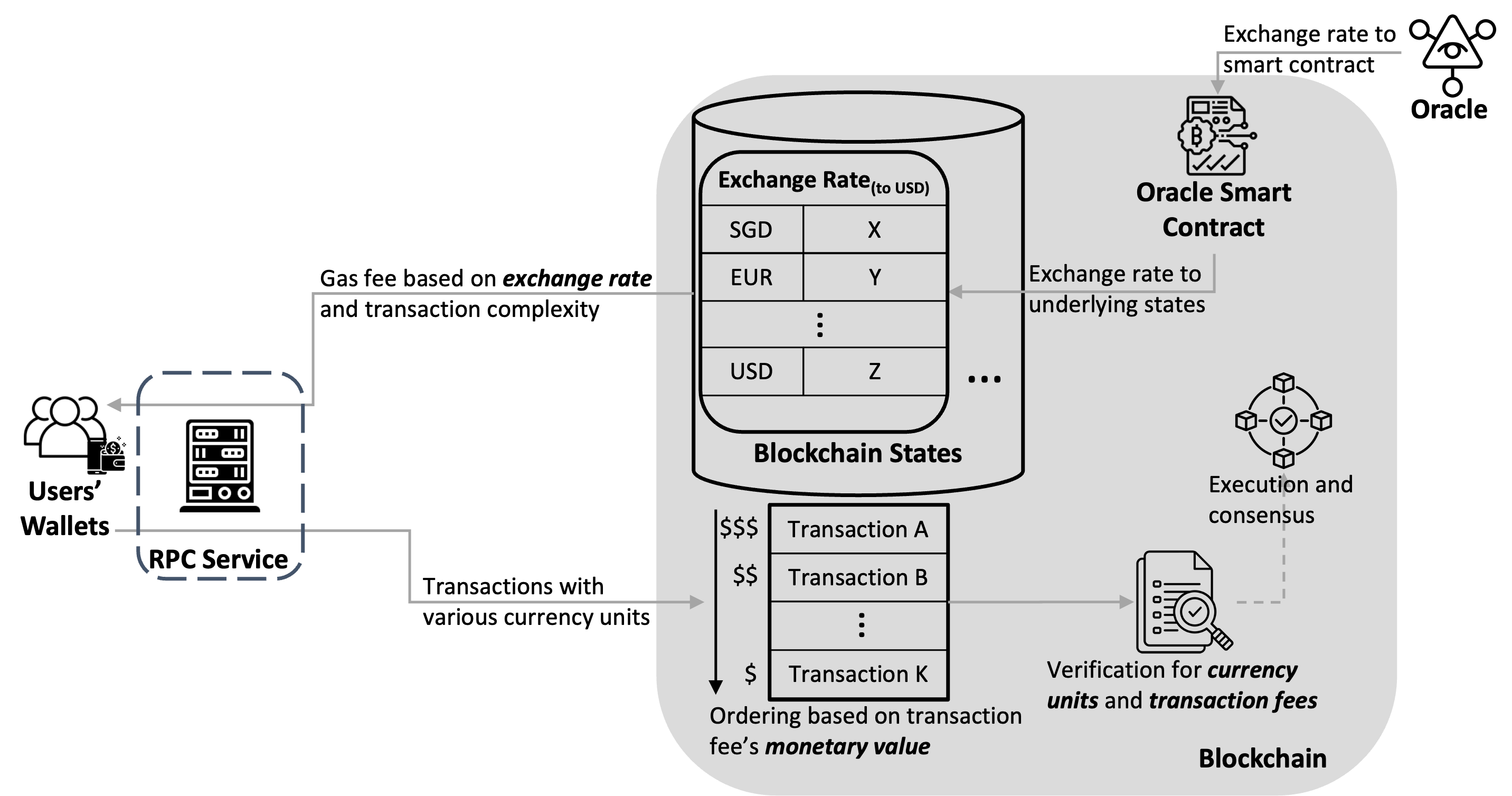}
\caption{Oracle-Based Gas Fee Adjustment. 
}
\label{fig:oracle}
\end{figure*}

Through the design presented in the previous subsection, we have largely achieved a balance between EVM compatibility and support for multiple native stablecoins. 
However, there is another important issue that needs to be resolved when it comes to paying transaction fees. 
That is, in order to ensure fairness, how to control the exchange rate between different stable coins, so that different users spend the same value when they choose different stable coins to settle the transaction fee.
To address this issue, we propose another core design: the Oracle-Based Gas Fee Adjustment.
The main idea of this design is illustrated in Figure \ref{fig:oracle}.

The core idea of this design is to obtain external exchange rate information through the oracle and feed this data to the system's base layer. 
Based on this exchange rate information, the system correspondingly adjusts the base fee of a transaction to ultimately control the exchange rates during transaction fee settlements. 
We proceed to elaborate on our design using intuition as a starting point.

\vspace{6pt}
\noindent
\textbf{Obtaining External Exchange Rate Information.}
To regulate the exchange rates during settlements, we first need to acquire exchange rate information from outside the blockchain, sourced from the real world. 
Our intuition is to leverage the oracle. 
To ensure the system's broad applicability, we decide to utilize existing oracle services to obtain external exchange rate information. 
However, existing oracle services typically can not provide data directly to the underlying system, but instead provide data to the smart contracts.
Considering this, we design and deploy a smart contract in the system to receive and store exchange rate data from the oracle. 
The oracle periodically invokes this smart contract to update its stored exchange rate data. 
This smart contract acts as an intermediary layer, relaying exchange rate data to the system's base layer without altering the existing oracle services.

Subsequently, to enable the system's base layer to retrieve and store exchange rate information, blockchain nodes periodically (e.g., every 10 blocks) fetch the exchange rate information from the smart contract based on the block height. 
The nodes then update and store this exchange rate information in the blockchain state they maintain.

\vspace{6pt}
\noindent
\textbf{Controlling Transaction Fee Exchange Rates.}
Once the system's base layer has obtained the exchange rate information, the next step is to actually control the exchange rates for transaction fees. 
Our approach is to regulate the exchange rates by controlling the base fee component of the transaction fee. 
As mentioned in Section \ref{background_other}, following the EIP-1559 upgrade, most EVM-compatible blockchain systems' transaction fees comprise a base fee and a tip. 
The base fee is determined by the system and cannot be altered by the user. 
Based on this consideration, our design dynamically links the base fee to the exchange rate information.
Whenever a blockchain node processes a transaction, it retrieves the stored exchange rate data corresponding to the transaction's currency unit from its maintained state. 
Subsequently, the node sets the transaction's base fee according to the current exchange rate data, thereby controlling the final transaction fee's exchange rate.

\vspace{3pt}
\noindent
\textit{Remarks.} 
There are several noteworthy points about the aforementioned design. 
First, our design ties the base fee to exchange rate information, a design that is different from Ethereum.
In Ethereum, the base fee adjusts according to network congestion, increasing when the network is congested. 
This is another factor, in addition to token price volatility, that affects the stability of transaction fees in blockchain systems like Ethereum. 
The purpose of MStableChain is to provide more stable transaction fees, and the above design is another means by which we can guarantee the stability of transaction fees in addition to stable coin settlement.
Second, in MStableChain, users can still set tips to flexibly increase their transaction fees, enabling faster transaction processing. 
Third, the total gas fee to be consumed by the end user for a transaction is still related to the complexity of that transaction (i.e., Equation \ref{equ:gas_fee}). 
The rules regarding gas usage (i.e., gas used) are consistent with Ethereum.
Finally, even though we change the way the base fee is calculated, the interaction with the blockchain remains EVM-compatible from the perspective of ordinary user.
This is because our design does not alter any parts of the EVM or the upper-layer applications.

\vspace{6pt}
\noindent
\textbf{Ensuring Transaction Ordering Unaffected by Exchange Rates.}
Our above design controls the exchange rate of transaction fees by adjusting the base fee. 
However, there remains a problem: when the system receives transactions from different users using different stablecoins to settle transaction fees, how to ensure that the ordering of transactions is not influenced by the different exchange rates of stablecoins.

For example, suppose User A chooses to settle the transaction fee with USD stablecoins, and the base fee at this time is 1 GWei (the unit of gas). 
Additionally, User A includes a tip of 1 GWei. 
Meanwhile, User B chooses to settle the transaction fee with CNY stablecoins and does not include a tip. 
According to our base fee adjustment mechanism, to ensure that the transaction fees paid by User A and User B are equivalent in value (excluding tips), the system sets the base fee for User B at 7.2 GWei (depending on the current exchange rate). 
Without additional mechanisms, the system would default to ordering transactions based on gas fee (similar to Ethereum). 
Consequently, User B's transaction would be processed before User A's transaction. 
However, this transaction ordering is unfair to User A, as its transaction fee has a higher monetary value (1 unit of USD base fee + 1 unit of USD tip), whereas User B's transaction fee only has a value equivalent to 1 unit of USD.

To address the aforementioned issue, we design a transaction ordering mechanism based on the value of the transaction fee. 
When a node receives transactions and needs to order transactions, it will retrieve the corresponding exchange rate data from its maintained state based on the transaction's currency unit. 
Subsequently, the node will use the current exchange rate data to calculate the \emph{real monetary value of the transaction fee}. 
Transactions are then ordered based on the value of the transaction fees, rather than the gas fee consumed, thereby eliminating the unfairness in transaction ordering caused by exchange rates.

With this mechanism, in the previous example, since User A's transaction fee has a higher monetary value, their transaction will be processed before User B's transaction.

\vspace{6pt}
\noindent
\textbf{Ensuring the Correctness of Transaction Fee Settlement.}
In a blockchain network, there may exist malicious nodes. 
It thus becomes a crucial issue to make the above mechanisms work properly even in the presence of malicious nodes.
Malicious nodes might alter the currency unit of a transaction, causing the transaction fee to be settled in an incorrect currency unit. 
For example, a user may choose to settle with a USD stablecoin, but a malicious node could change it to a EUR stablecoin during transaction processing.

To address this issue, we incorporate a comparative verification of transaction currency units and their transaction fees during transaction processing.
When a node validates a transaction, it must additionally verify that the exchange rate of the base fee corresponds to the currency unit of the transaction. 
If any mismatch is detected, the transaction is deemed invalid. 
Eventually, the consistency of transaction processing across all nodes is then guaranteed by consensus, thereby excluding the influence of malicious nodes.

\subsection{On-Chain Voting-Based Native Stablecoin Management}
\label{design_mgmt}

The aforementioned designs basically meet the needs of ordinary users for settling transaction fees with multiple native stablecoins when interacting with MStableChain. 
However, for entities backing the stablecoins, an important issue remains unresolved: how to manage the native stablecoins. 
To address this issue, we propose another key design: a series of On-Chain Voting-Based Native Stablecoin Management mechanisms.

Unlike typical cryptocurrencies (e.g., ETH), stablecoins often require additional management by the entities backing them. 
However, in MStableChain, the stablecoins are native stablecoins rather than smart contract-based stablecoins.
As a result, it is difficult to manage the native stablecoins in our system through smart contracts.
Therefore, we design a series of native stablecoin management functions at the system level, allowing relevant entities to manage native stablecoins through proposal and voting transactions.

Specifically, the main management functions for native stablecoins in MStableChain include \emph{mint, burn, whitelist, blacklist, committee size, and quorum size}. 
The mint function is primarily responsible for minting a corresponding number of stablecoins on the chain upon receipt of a certain amount of fiat currency by the entity endorsing a particular stablecoin.
The burn function, conversely, destroys a corresponding amount of stablecoins on the chain when fiat currency is withdrawn from the entity backing the stablecoin. 
The whitelist function specifies which individuals (addresses) are eligible to manage a particular stablecoin.
The blacklist function specifies those who are not allowed to use a stablecoin. 
The blacklist design is primarily for regulatory purposes, allowing entities to ban users engaged in misconduct (e.g., money laundering). 
The committee size function is primarily responsible for regulating the size of the committee that manages a stablecoin. 
The intuition behind this design is that, for management security reasons, an existing entity endorsing a stablecoin may need to have multiple people working together to manage the stablecoin, such as multiple members of the management team of a particular entity.
The quorum size function is responsible for regulating how many people must be in agreement before a management for a stablecoin can be passed. 
For example, if the committee size is 4 and the quorum size is 3, that means that at least 3 out of the 4 people must agree on a management for it to take effect.

To be able to perform the various functions described above, we design proposal transactions and voting transactions.
These transactions are implemented at the system level and are native transactions to MStableChain, similar to simple money transfer transactions (not smart contract transactions). 
To initiate a certain management (e.g., minting) of a stablecoin, the entity endorsing it (or an individual within the entity) first initiates a proposal transaction (e.g., a minting proposal) to propose the management action.
That transaction is broadcast, consensused, and recorded on the blockchain. 
As previously stated, there may be more than one person in a given entity that has management over a stablecoin.
Therefore, the individuals responsible for managing the stablecoin in that entity can submit voting transactions to vote on the proposal transaction, determining whether the proposal is accepted. 
Similarly, voting transactions are broadcast, reach consensus, and are recorded on the blockchain to ensure security. 
Additionally, for the sake of management transparency, the relevant entity may attach additional proof information to the data fields of the transaction (e.g., evidence related to money laundering by a particular misbehaving user). 
This supporting information could be the data itself or a URL pointing to a particular piece of data.

\vspace{3pt}
\noindent
\textit{Remarks.} 
Given that MStableChain supports multiple native stablecoins, the aforementioned design is independent for each native stablecoin. 
That is to say, each primary stablecoin may be backed by an entity that can utilize the above design to perform management functions for its stablecoin.
The management of one native stablecoin does not affect other native stablecoins. 
This is because we maintain the state of different native stablecoins independently at the bottom of the system.

\subsection{Interaction Process in MStableChain}
\label{design_workflow}

We now provide an example of the interaction process with MStableChain to help readers better understand the overall system workflow. 

Suppose a user interacting with a dApp wishes to pay the transaction fee using the USD stablecoin. 
The user would first switch the RPC in their wallet to select the RPC corresponding to the USD stablecoin.
At this point, the corresponding RPC nodes retrieve the balance state of the corresponding currency unit (USD) under the user's account from the blockchain and displays it in the user's wallet interface. 

When the user initiates a transaction, the wallet needs to determine the transaction fee via RPC communication. 
The oracle continuously feeds exchange rates for various stablecoins from external sources to the underlying state of the blockchain via the smart contract.
The base fees of different native stablecoins are then dynamically adjusted according to the exchange rate.
As a result, the RPC nodes can retrieve the gas fee corresponding to the USD stablecoin for this transaction from the blockchain and display it in the user's wallet interface. 

When a user's transaction is sent to the blockchain network, it undergoes a process where it is ordered, validated, executed, and recorded alongside other transactions through the consensus process. 
While these transaction processing steps are similar to those in most existing systems, there are a few noteworthy points in MStableChain. 
First, transactions are ordered based on the actual monetary value of the transaction fees, rather than just the nominal gas fee. 
Second, when validating transactions, each node must additionally verify the currency unit and the corresponding base fee to ensure consistency and correctness.

\section{Analysis and Discussion}

\subsection{Security Analysis}
\label{analysis_security}

We now analyze the security of MStableChain.

\begin{theorem}
MStableChain is secure when the proportion of malicious blockchain nodes within the system is less than 1/3.
\end{theorem}

\begin{proof}
First, regarding the consensus protocol, MStableChain utilizes the Tendermint consensus protocol. 
Tendermint is a Byzantine Fault Tolerant (BFT) consensus protocol that ensures both safety and liveness (i.e., security) as long as the proportion of malicious nodes remains below 1/3. 
In MStableChain, blockchain nodes are responsible for running the consensus protocol. 
Therefore, as long as the proportion of malicious blockchain nodes within the MStableChain system is less than 1/3, the consensus protocol remains secure.

Second, for the multi-currency units, multi-type RPCs, and the oracle-based gas fee adjustment mechanisms, malicious blockchain nodes may attempt to tamper with the currency unit or gas fee of a transaction to force users to settle using incorrect currencies or fees. 
However, due to the protection provided by the consensus protocol, such attacks cannot succeed in MStableChain.

Specifically, when a user initiates a transaction, its wallet retrieves gas fee-related information from the blockchain. 
In MStableChain, exchange rate information is obtained through honest oracles (as detailed in the threat model in Section \ref{model_threat}) and determines the user's base fee. 
Malicious blockchain nodes might try to manipulate the returned base fee to make users settle transactions with fees that do not align with the correct exchange rates. 
However, when the user's transaction is broadcast to the blockchain via honest RPCs (as detailed in the threat model in Section \ref{model_threat}), during the consensus process, the consensus protocol will additionally verify whether each transaction's currency unit and exchange rate information match (as discussed in Section \ref{design_oracle}). 
If any mismatches are detected, the blockchain nodes will determine the transaction to be invalid through the consensus protocol.

Moreover, malicious blockchain nodes might attempt to locally alter the currency unit of a transaction. 
However, since transactions are broadcast across the entire network, malicious blockchain nodes cannot alter the transaction on a network-wide scale. 
Therefore, during the consensus process, honest blockchain nodes will detect inconsistencies if multiple versions of the same transaction exist. 
As long as the proportion of malicious blockchain nodes within the system is less than 1/3, the consensus protocol ensures that these tampered transactions cannot be accepted.

Finally, regarding the on-chain native stablecoin management mechanism, each management function is implemented as a specific type of transaction. 
When the relevant entities need to perform a particular management task, they send the corresponding transaction to the blockchain. 
The blockchain nodes in MStableChain will then validate the transaction's content and use consensus to ensure its validity. 
Given that the consensus protocol remains secure as long as the proportion of malicious blockchain nodes is less than 1/3, the management of native stablecoins is also secure under these conditions.

\end{proof}

\subsection{Handling Stablecoin Fees}
\label{analysis_fee}

How MStableChain handles stablecoin transaction fees paid by users is a topic worth discussing. 
In some existing blockchain systems (e.g., Ethereum), a portion of the transaction fees paid by users is burned by the system. 
However, MStableChain uses stablecoins for transaction fee settlements. 
Since stablecoins are pegged to fiat currencies, MStableChain cannot arbitrarily burn the collected stablecoin transaction fees. 
For this reason, the stablecoin transaction fees paid by users in MStableChain are distributed to the blockchain nodes rather than being burned. 
This approach ensures that the stablecoins on the chain remain pegged to the fiat currencies managed by the corresponding off-chain entities. 
Additionally, it incentivizes blockchain nodes to actively and correctly process transactions.


\subsection{Adding and Removing Stablecoins}
\label{analysis_add}

MStableChain supports the dynamic addition and removal of native stablecoin types within the system to cater to more flexible and broader application scenarios. 
When an entity wishes to add or remove a type of native stablecoin that it backs in MStableChain, it must first submit an application to the project team. 
Subsequently, MStableChain will upgrade the system to correspondingly add or remove the relevant stablecoin information (e.g., balance state, currency unit state) at the system level. 
After this, the entity can govern its stablecoin through the on-chain native stablecoin management mechanisms we design.

\subsection{Generality}
\label{analysis_generality}

MStableChain possesses excellent general applicability. 
With some modifications, MStableChain can be adapted for consortium blockchains or scenarios with strict compliance requirements. 
For example, by utilizing the concept of multiple native stablecoins for transaction fee settlement in MStableChain, a compliance-focused blockchain system can employ institutions like central banks to manage digital fiat currencies (e.g., Central Bank Digital Currency (CBDC) \cite{islam2022privacy}) on the chain while supporting multiple digital fiat currencies for transaction fee settlement simultaneously. 
Since our approach does not involve issuing any cryptocurrencies, it helps enhance compliance.

Additionally, the concept of multiple native stablecoins for transaction fee settlement in MStableChain can also address the incentive issues in consortium blockchains. 
Existing consortium blockchains are typically run voluntarily by member institutions. 
However, this non-incentive-based operation significantly hinders the widespread adoption of consortium blockchain applications. 
By adopting MStableChain's concept, consortium blockchains can incentivize nodes using stablecoins or digital fiat currencies without issuing any cryptocurrencies. 
This represents a potential future research direction for us.

\subsection{Limitations}
\label{analysis_limitation}

Despite the numerous benefits MStableChain offers, we acknowledge certain potential limitations. 
For example, as previously mentioned, adding or removing native stablecoins requires a system upgrade, which can be somewhat cumbersome. 
However, we have designed mechanisms to mitigate this issue. 
First, we design and implement such system upgrades as seamless upgrades.
This means that the blockchain network does not require any soft or hard forks, nor do blockchain nodes need to delete and reload their recorded states. 
Instead, blockchain nodes only need to download the latest blockchain client binary files before a specified time point (usually at a designated block height). 
Once the designated time point is reached, the blockchain nodes will automatically switch to running the latest blockchain client. 
This process takes only a few seconds. 
Second, we develop a specific SDK to facilitate quick and easy code modifications for adding or removing native stablecoins.

Another potential limitation is that the management mechanism for native stablecoins in MStableChain is implemented through different types of native transactions at the system level, which are not EVM-compatible. 
This means that relevant entities cannot directly use existing EVM-compatible applications to manage native stablecoins in MStableChain. 
To address this issue, we design specialized applications (for both desktop and mobile) for managing native stablecoins on MStableChain. 
Relevant entities can use these applications directly or customize them for more convenient management of native stablecoins. 
It is worth noting that the entities managing native stablecoins typically have specific managerial roles, while MStableChain continues to provide EVM compatibility for general users.

\section{Implementation and Evaluation}
\label{sec:evaluation}

\subsection{Implementation}
\label{subsec:implementation}

We primarily implement the MStableChain prototype using Go and Rust. 
Without loss of generality, the prototype system supports 2 different native stablecoins: 
one based on the US dollar (USD) and another based on the Chinese yuan (CNY). 
For the consensus, we utilize the Tendermint consensus protocol, which was developed in Go. 
We adopt the Rust version of the EVM \cite{rustevm} and develop the runtime based on it for transaction processing. 
Additionally, the RPC is developed following the Ethereum RPC standard, and the oracle smart contracts are written in Solidity, with Chainlink chosen as the oracle. 
Furthermore, as mentioned in Section \ref{analysis_limitation}, for ease of use, we also provide a set of SDKs for developers and dedicated applications for stablecoin managers. 
A Beta version of the blockchain system based on MStableChain has been launched on its Mainnet \cite{hela}.

For comparative evaluation, we further implement a prototype of the Gas Station Network (GSN) protocol as a baseline. 
As described in Section \ref{background_volatile}, GSN is a smart contract-based protocol that offers users multiple non-native tokens as payment options. 
However, it has drawbacks such as high overhead and increased latency, which we will demonstrate in the subsequent experiments.
Moreover, we also build an Ethereum local network to use as a comparison.

\subsection{Experimental Setup}
\label{subsec:setup}

We set up an MStableChain blockchain test network locally using several servers. 
Four servers are used to run the blockchain nodes. 
Two additional servers are dedicated to running the RPC services: 
One for the RPC node corresponding to the USD stablecoin and the other for the RPC node corresponding to the CNY stablecoin. 
Another server is used to run the oracle service. 
Each server is with 8-core CPUs and 16G memory.
To ensure fairness in evaluation, we adjust MStableChain's block production time to be consistent with Ethereum, setting it to 12 seconds. 
Additionally, we maintain consistency with Ethereum in terms of the transaction size and how gas usage is calculated (e.g., more complex transactions consume more gas).

\begin{figure}[t]
\subfloat[]{\includegraphics[scale=0.345]{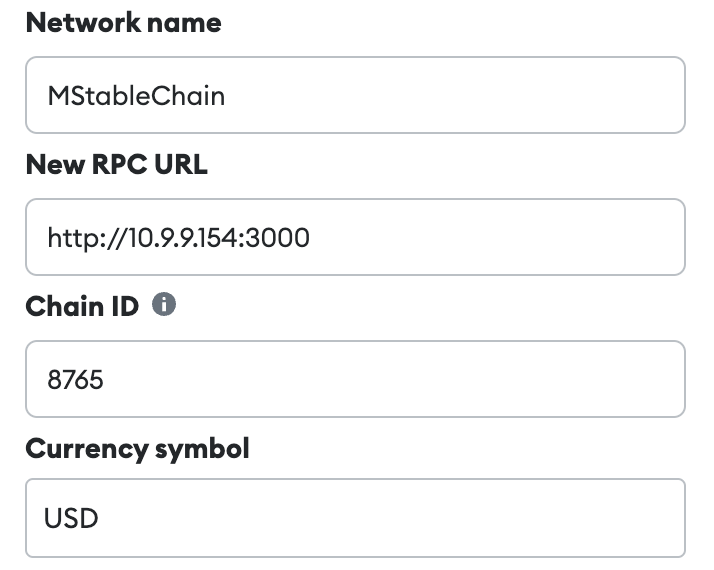}\label{subfig:RPCUSD}}
\hfill
\subfloat[]{\includegraphics[scale=0.345]{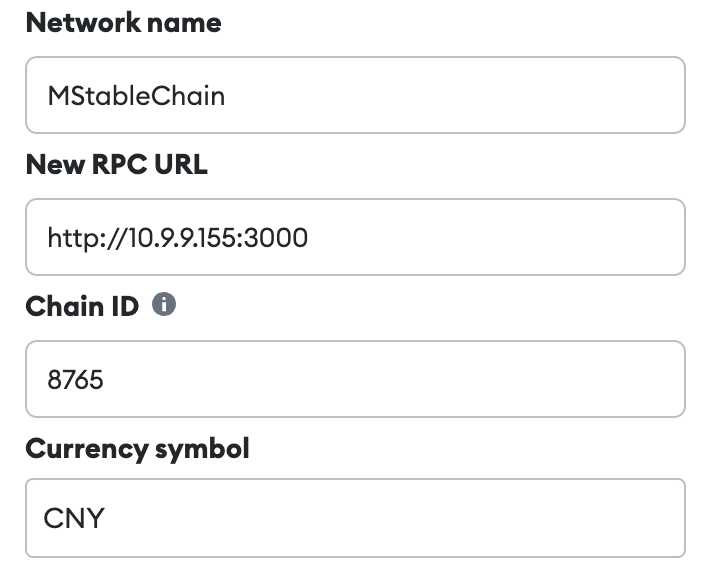}
\label{subfig:RPCCNY}
}
\caption{User-configured URLs for different RPCs. (a) RPC corresponding to USD. (b) RPC corresponding to CNY.}
\label{fig:RPC}
\end{figure}

\begin{figure}[t]
\subfloat[]{\includegraphics[scale=0.51]{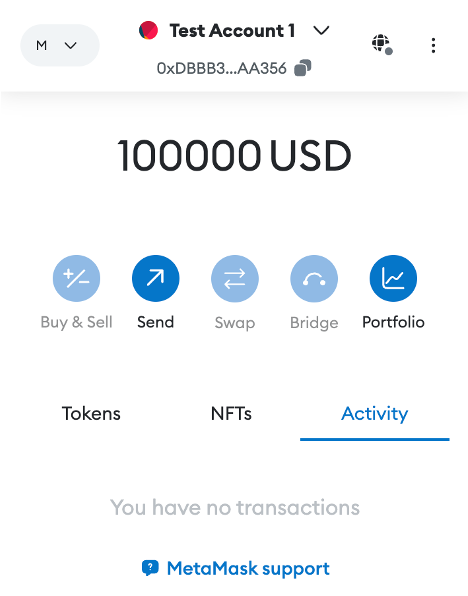}\label{subfig:USD}}
\hfill
\subfloat[]{\includegraphics[scale=0.51]{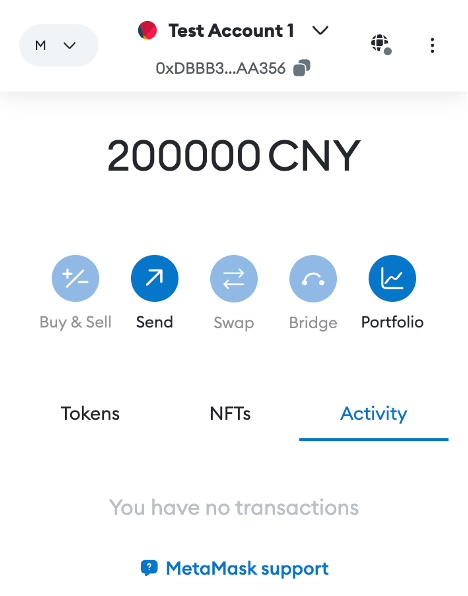}
\label{subfig:CNY}
}
\caption{The user's wallet interface after connecting different RPCs. (a) Interface when connecting the RPC corresponding to USD. (b) Interface when connecting the RPC corresponding to CNY.}
\label{fig:Interface}
\end{figure}

\subsection{User Interaction with Multi-Native Stablecoins}
\label{subsec:interaction}

In this part, we will demonstrate how users can interact with multiple native stablecoins in MStableChain using an EVM-compatible wallet. 
We use MetaMask, the most well-known EVM-compatible wallet, as an example to illustrate the process. 
As shown in Figure \ref{fig:RPC}, to interact with multiple native stablecoins in MStableChain, users need to configure 2 different RPC URLs when selecting the RPC to connect to. 
The URL http://10.9.9.154:3000 corresponds to the RPC for the USD stablecoin, while the URL http://10.9.9.155:3000 corresponds to the RPC for the CNY stablecoin.

As depicted in Figure \ref{fig:Interface}, after successfully connecting to the respective RPC, the user can switch between the stablecoins to view their corresponding balances in the wallet. 
For instance, in our test account, there are 100,000 USD stablecoins and 200,000 CNY stablecoins. 
When connected to a particular RPC, the user can send various transactions through that RPC and use the corresponding stablecoin to settle transaction fees.

\begin{figure}[t]
\subfloat[]{\includegraphics[scale=0.51]{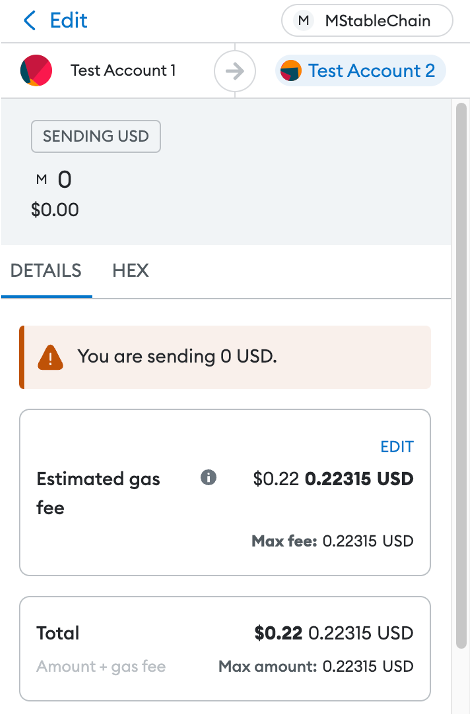}\label{subfig:feeUSD}}
\hfill
\subfloat[]{\includegraphics[scale=0.51]{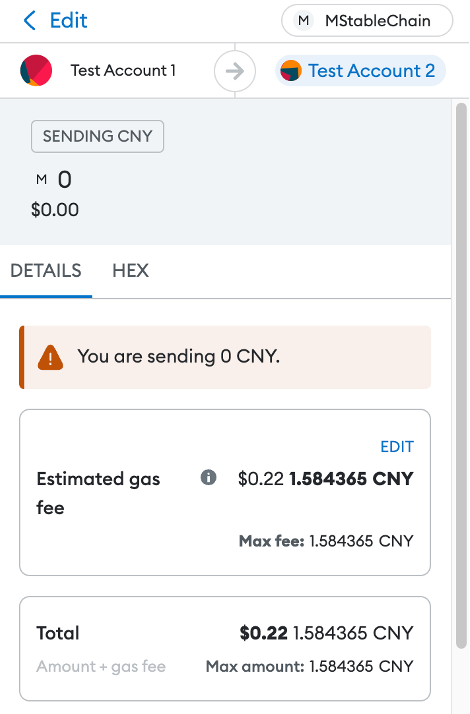}
\label{subfig:feeCNY}
}
\caption{Transaction fee when paying in different stablecoins (determined by the exchange rate). (a) Fees when paying with USD stablecoin. (b) Fees when paying with CNY stablecoin.}
\label{fig:fee}
\end{figure}

\subsection{Transaction Fees Controlled by Exchange Rate}
\label{subsec:txn_fee}

We now demonstrate how transaction fees are controlled by exchange rates. 
We use a test user to send the same transaction into MStableChain using USD stablecoin and CNY stablecoin as settlement respectively, and observe the consumption of its transaction fee. 
To show more clearly the effect of exchange rate on the transaction fee, we set the transaction as a transaction with a transfer amount of 0. 
In this way, the cost during the transaction will be entirely due to the transaction fee.
The results are shown in Figure \ref{fig:fee}.

When the user chooses to settle the transaction fees with USD stablecoins, the total transaction fee is 0.22315 USD.
When the user opts to use CNY stablecoins for the transaction fee settlement, the total transaction fee consumed is 1.584365 CNY, which is equivalent to 0.22 USD.
It is important to note that the user does not set a tip when sending the transaction, so the total gas fee is calculated as the base fee multiplied by the gas consumption (i.e., gas used). 
Since the same transactions are sent, the gas consumption remains constant. 
Therefore, we can observe that the total gas fee consumed when using CNY stablecoins for settlement is 7.11 times that of using USD stablecoins, 
which aligns with the exchange rate between the two currencies at the time of evaluation.

\begin{figure}[t]
\centering
\includegraphics[scale=0.33]{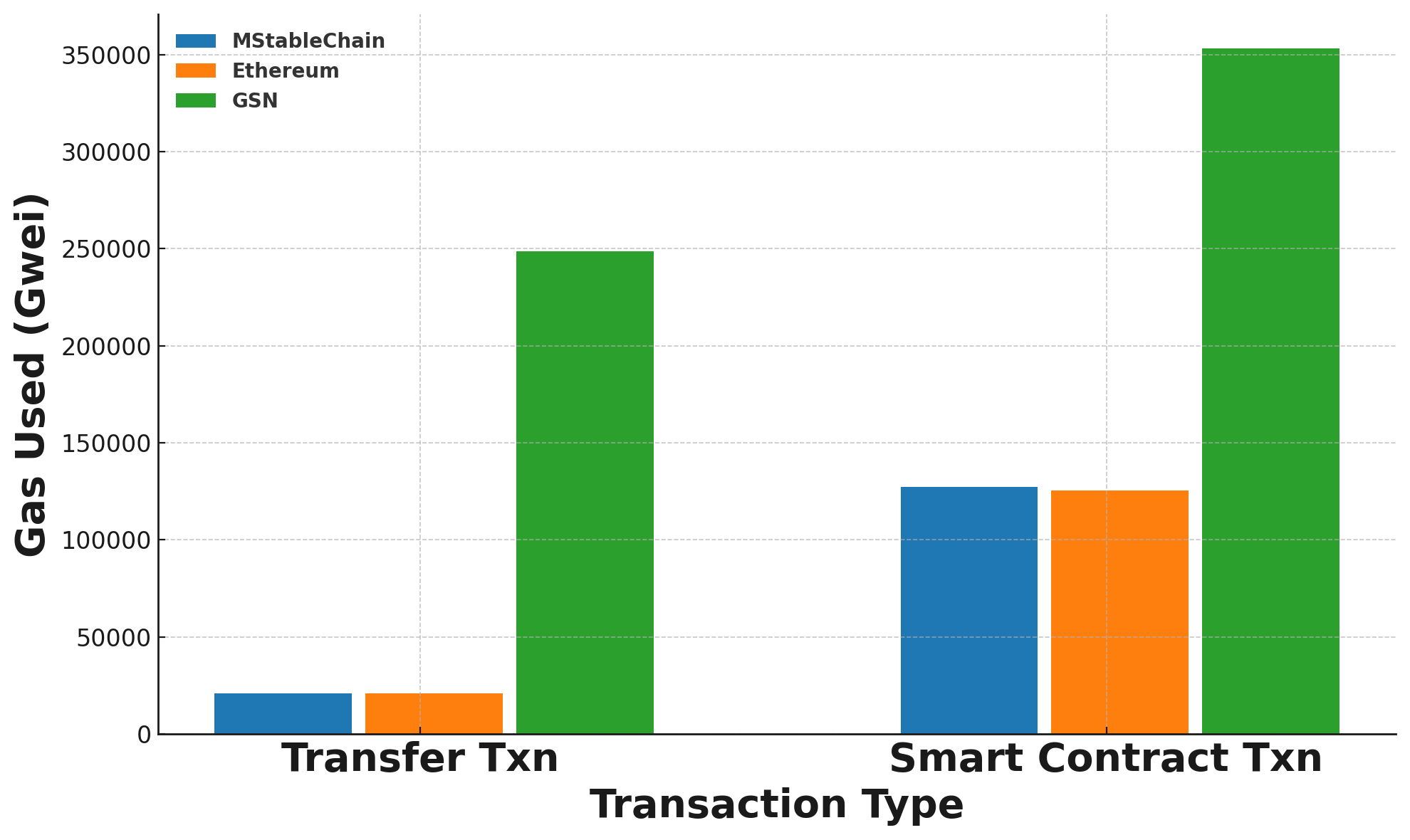}
\caption{Comparative results of gas consumption. 
}
\label{fig:gas_used}
\end{figure}

\subsection{Gas Consumption and Latency}
\label{subsec:gas_consumtion}

We now compare MStableChain with GSN and Ethereum to evaluate and contrast gas consumption (i.e., Gas Used in Equation \ref{equ:gas_fee}) and transaction latency. 
We send 10 standard transfer transactions and 10 simple smart contract transactions to MStableChain, the GSN protocol, and the local Ethereum network, respectively, and record the average gas consumption and transaction latency for each.

As shown in Figure \ref{fig:gas_used}, when sending transfer transactions and smart contract transactions, MStableChain consumes 21,000 and 127,236 units of gas, respectively, which is similar to the Ethereum network. 
In contrast, GSN consumes 248,612 and 353,068 units of gas for transfer transactions and smart contract transactions, respectively. 
Compared to GSN, the gas consumption of MStableChain is only 8.4\% and 36\% of that of GSN, respectively.
This significant difference is primarily due to MStableChain's use of multiple native tokens for transaction fee settlements. 
This approach eliminates the need for the system to execute additional complex smart contract logic when processing transactions, thereby avoiding extra gas consumption.
It is also important to note that MStableChain's oracle price feeds run independently on a periodic basis, so they do not impact the gas consumption of user transactions. 
On the other hand, GSN includes numerous smart contracts, which require executing a substantial amount of complex smart contract logic for each transaction. 
This complexity significantly increases gas consumption for transactions processed through GSN.

\begin{figure}[t]
\centering
\includegraphics[scale=0.33]{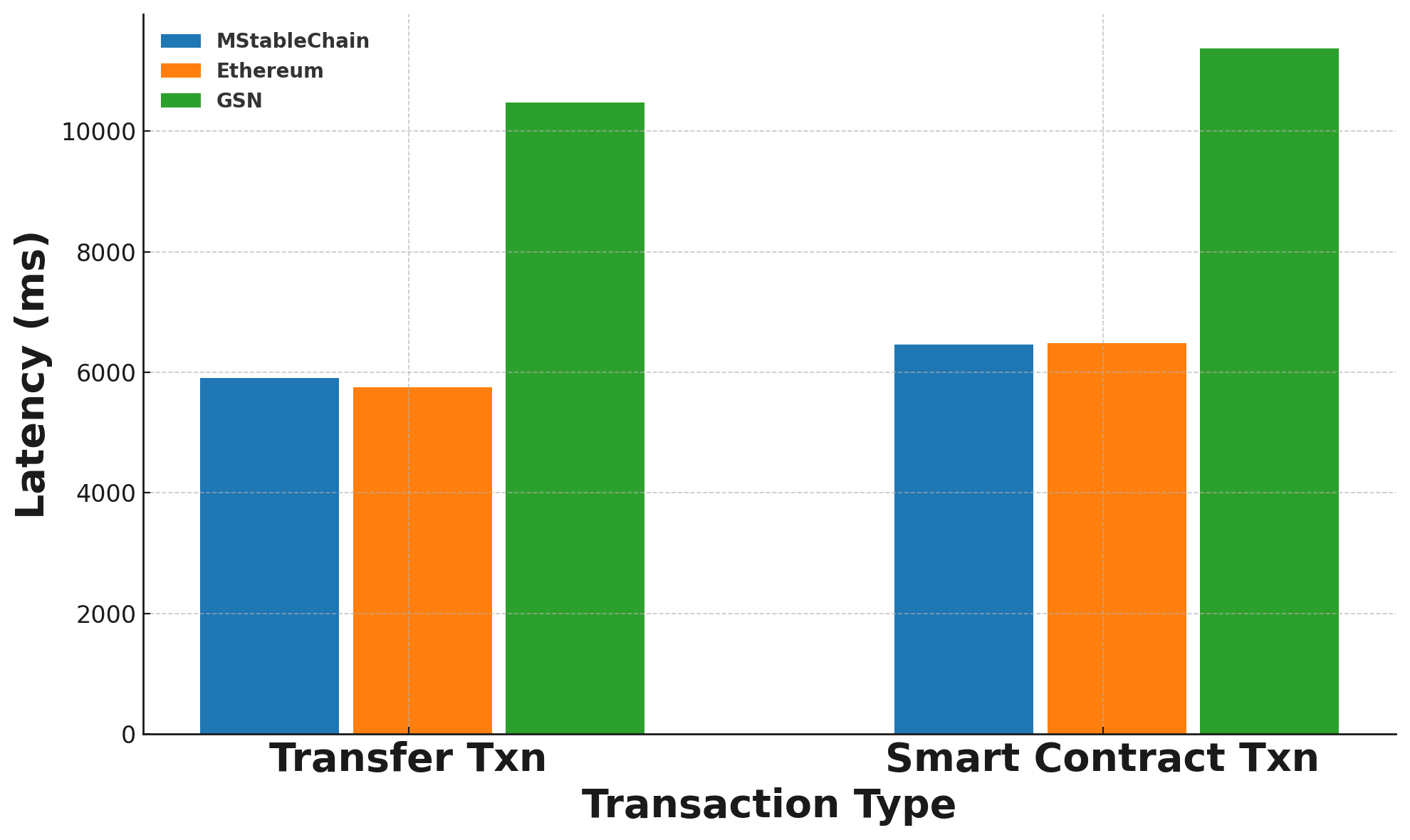}
\caption{Comparative results of transaction latency. 
}
\label{fig:latency}
\end{figure}

As shown in Figure \ref{fig:latency}, the transaction latency for sending transfer transactions and smart contract transactions on MStableChain is 5,903 ms and 6,467 ms, respectively, which is similar to the Ethereum network. 
In contrast, the transaction latency for GSN is 10,478 ms and 11,373 ms, respectively. 
Compared to GSN, MStableChain's transaction latency is only 56.3\% and 56.9\% of that of GSN, respectively.
This is primarily because MStableChain's critical transaction processing path is similar to that of Ethereum, with minimal additional computational and I/O overhead (on the order of milliseconds). 
The oracle price feeds required by MStableChain execute independently on a periodic basis and are outside the critical transaction processing path, so they do not affect transaction processing latency.
On the other hand, the GSN protocol includes several additional components, such as relayers, which are part of the critical transaction processing path. 
These additional components significantly increase the transaction processing latency in GSN.

\subsection{Transaction Fee Stability}
\label{subsec:stability}

\begin{figure}[t]
\centering
\includegraphics[scale=0.33]{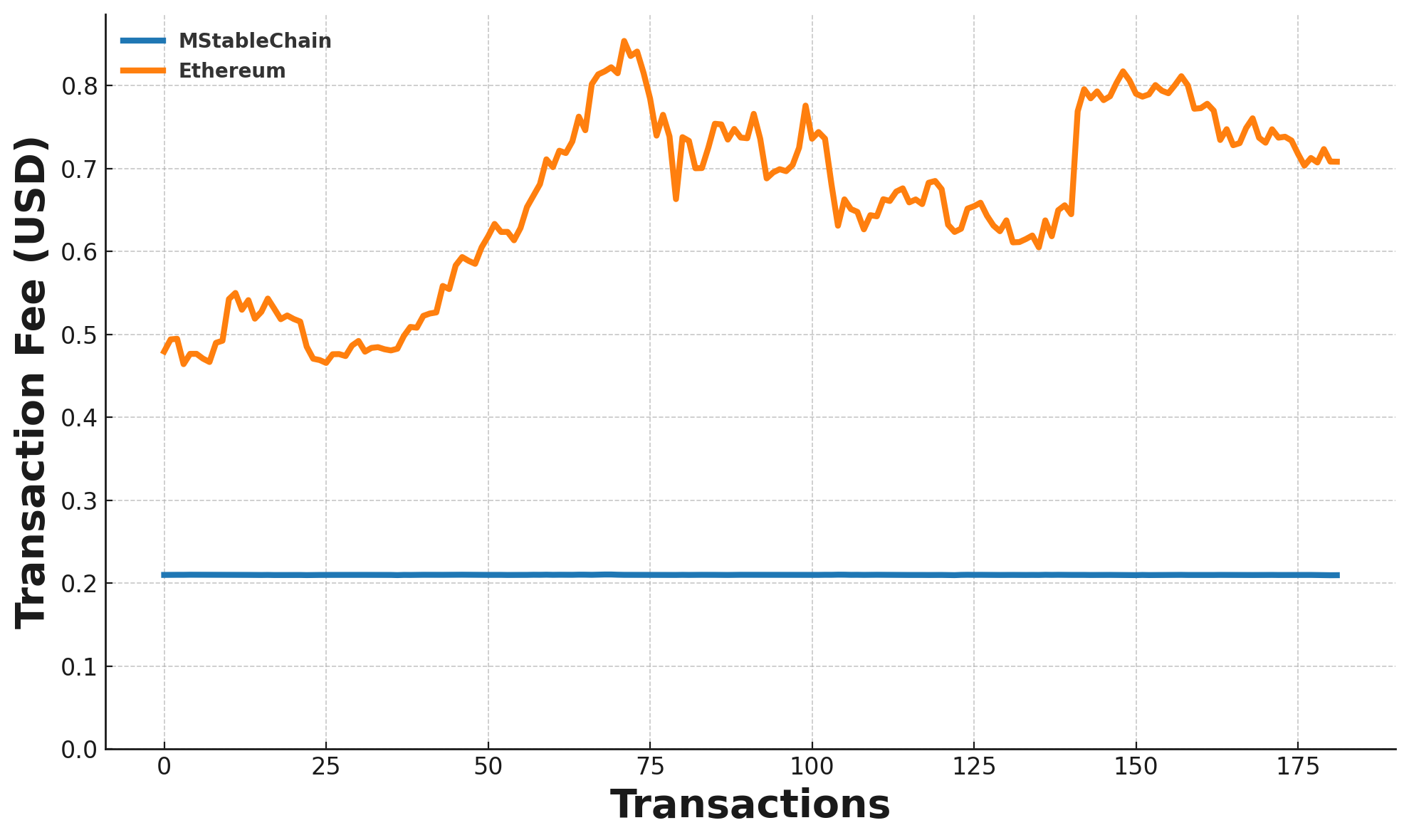}
\caption{Fluctuations in transaction fee prices. 
}
\label{fig:price_stability}
\end{figure}

We evaluate the stability of transaction fees in MStableChain now. 
We send transfer transactions to both the local Ethereum network and MStableChain and calculate the transaction fees based on the daily prices of Ethereum and USDT (a USD stablecoin) from January 1 to June 30, 2024. 
The results are shown in Figure \ref{fig:price_stability}. 
As can be seen, due to the volatility of Ethereum's token price, its transaction fees also exhibits significant fluctuations, with the highest fee being 1.840 times greater than the lowest. 
In contrast, MStableChain, which uses stablecoins for transaction fee settlements, shows almost no fluctuation in transaction fees. 
The difference between the highest and lowest fees is only 1.004 times.

\section{Conclusions}

In this paper, we present MStableChain, an innovative blockchain system that aims at addressing the prevalent issues of transaction fee volatility and inflexibility in payment methods in existing blockchain systems. By incorporating multiple stablecoins pegged to fiat currencies as native tokens, the system successfully achieves stable transaction fees and a user-friendly payment experience. MStableChain supports the settlement of transaction fees in multiple native stablecoins without altering EVM compatibility for mass adoption through its multi-currency units and multi-type RPCs mechanism. Additionally, the oracle-based gas fee adjustment mechanism ensures fairness in transaction fees across different stablecoins, and the introduction of an on-chain voting-based native stablecoin management mechanism enhances the system's security and transparency. Experimental results demonstrate that MStableChain offers significant advantages in terms of usability and fee stability, providing a practical path toward broader adoption in the blockchain ecosystem.

\bibliographystyle{IEEEtran}
\bibliography{ref}

\vspace{-30pt}
\begin{IEEEbiography}[{\includegraphics[width=1in,height=1.25in,clip,keepaspectratio]{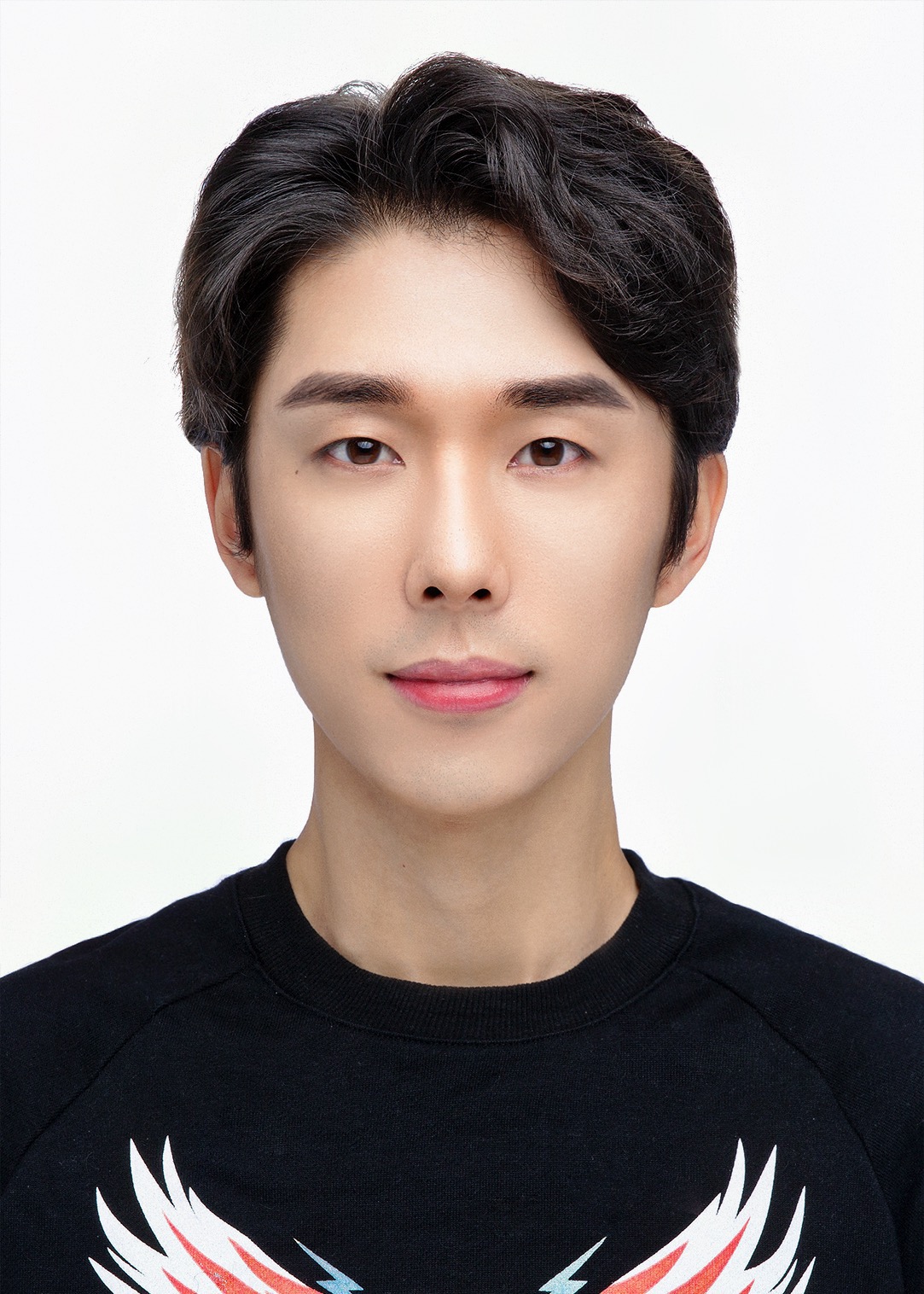}}]{Mingzhe Li}
is currently a Scientist with the Institute of High Performance Computing (IHPC), A*STAR, Singapore.
He received his Ph.D. degree from the Department of Computer Science and Engineering, Hong Kong University of Science and Technology in 2022.
Prior to that, he received his B.E. degree from Southern University of Science and Technology.
His research interests are mainly in AI for blockchain, blockchain interoperability, blockchain security, sharding-based blockchain, Web 3.0, DeFi, network economics, etc.
\end{IEEEbiography}
\vspace{-30pt}
\begin{IEEEbiography}
[{\includegraphics[width=1in,height=1.25in,clip,keepaspectratio]{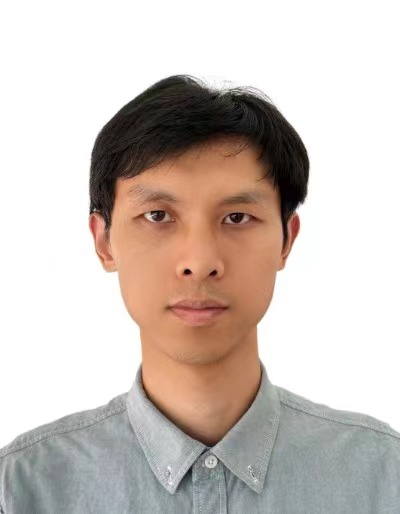}}]{Bo Gao}
received his Ph.D. degree from SUTD in 2022, and joined ASTAR thereafter. He obtained his B.Sc. and M.Sc. from Xi’an Jiaotong University, China in 2012 and 2014. He worked as a senior Industrial Engineer in Eaton (China) Investments Co., Ltd. until 2016. 
He now works in IHPC, and his research interests include formal methods, symbolic execution, blockchain and smart contracts.
\end{IEEEbiography}
\vspace{-30pt}
\begin{IEEEbiography}[{\includegraphics[width=1in,height=1.25in,clip,keepaspectratio]{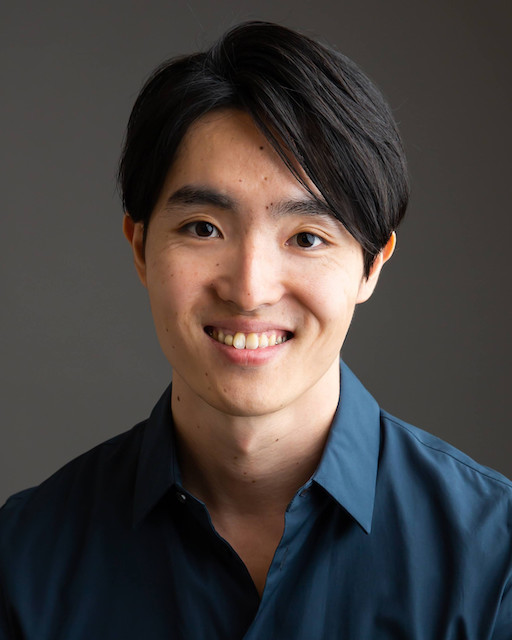}}]{Kentaroh Toyoda}
was born in Tokyo, Japan in 1988. He received B.E., M.E., and Ph.D. (Engineering) degrees in the Department of Information and Computer Science, Keio University, Yokohama, Japan, in 2011, 2013, and 2016, respectively. He was an assistant professor at Keio University from Apr. 2016 to Mar. 2019 and is a senior scientist at the Institute of High Performance Computing (IHPC), Agency for Science, Technology and Research (A*STAR), Singapore. His research interests include blockchain, AI, mechanism design, security and privacy.
\end{IEEEbiography}
\vspace{-30pt}
\begin{IEEEbiography}
[{\includegraphics[width=1in,height=1.25in,clip,keepaspectratio]{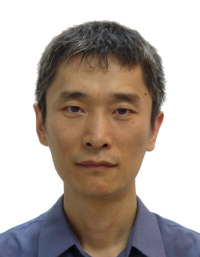}}]{Yechao Yang}
is holding a Bachelor degree from the University of Science and Technology of China. He was working on different domains including Windows/Linux Device Drivers, UI/UX with QT, Web Application backend and Distributed System. His preferred programming languages include C/C++, Java, JavaScript, Python.
\end{IEEEbiography}
\vspace{-30pt}
\begin{IEEEbiography}
[{\includegraphics[width=1in,height=1.25in,clip,keepaspectratio]{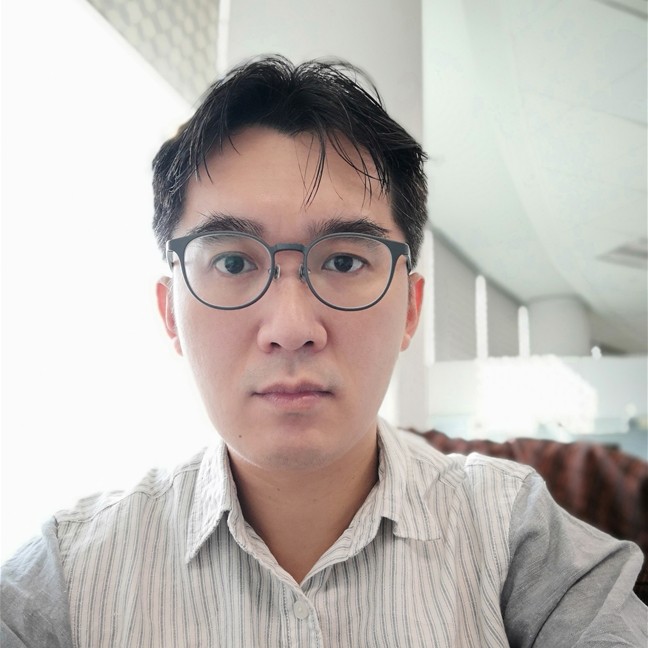}}]{Juniarto Samsudin}
is currently a Lead Research Engineer at A*STAR's Institute of High Performance Computing. He holds a Master's degree in Smart Product Design from Nanyang Technological University, Singapore. His research interests include Federated Learning, Web3, IoT, and Linux firmware. He enjoys running, playing music and singing on the weekend.
\end{IEEEbiography}
\vspace{-30pt}
\begin{IEEEbiography}
[{\includegraphics[width=1in,height=1.25in,clip,keepaspectratio]{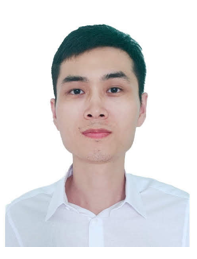}}]{Haibin Zhang}
received Bachelor degrees in Computer Science and Animal Science in 2010. He has been actively involved and passionate about the fields of AI and blockchain for years, having worked as a solution engineer and led research and development teams at NEC Corp and Alibaba Group from 2018 to 2020. He is currently a research engineer at the Institute of High Performance Computing, A*STAR Singapore, with research interests in federated learning, blockchain, high-performance computing, and distributed storage systems. 
\end{IEEEbiography}
\vspace{-30pt}
\begin{IEEEbiography}
[{\includegraphics[width=1in,height=1.25in,clip,keepaspectratio]{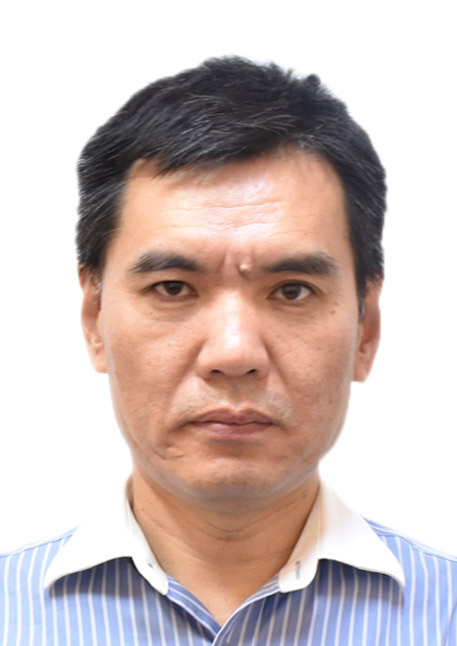}}]{Sifei Lu}
received his Bachelor's and Master's degrees from Tsinghua University in Beijing, China, in 1990 and 1995, respectively. He is currently a Principal Research Engineer at the Institute of High Performance Computing (IHPC), A*STAR, Singapore. His research interests include data analysis, AI models, and blockchain systems.
\end{IEEEbiography}
\vspace{-30pt}
\begin{IEEEbiography}
[{\includegraphics[width=1in,height=1.25in,clip,keepaspectratio]{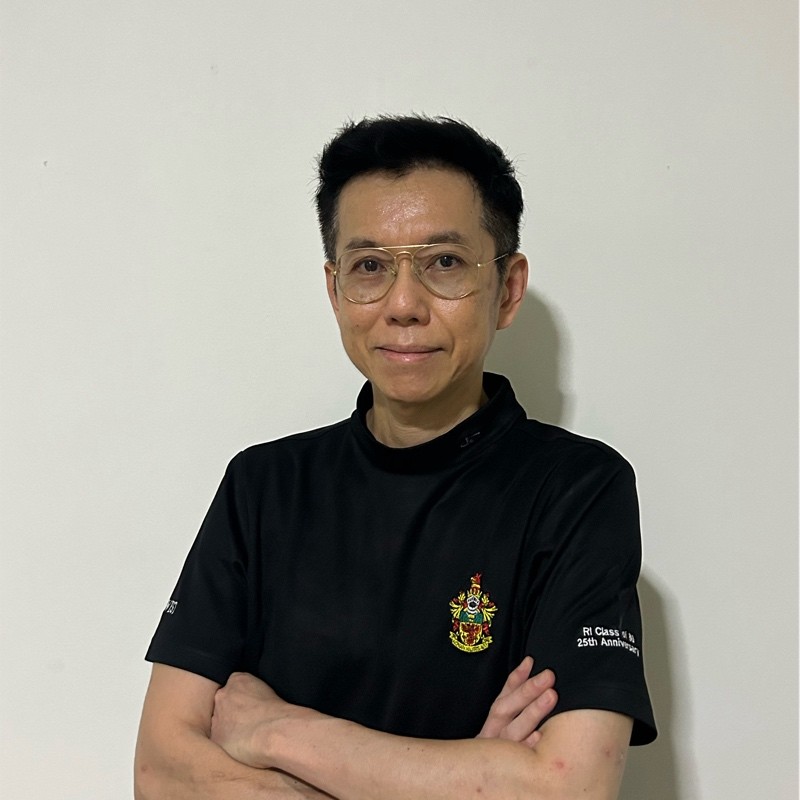}}]{Tai Hou Tng} is currently an innovation lead with the Institute of High Performance Computing, A*STAR. He leads the
design of innovation solution through working with industry partners. Highly skilled
engineer with various experience in software development and machine learning.
His research interest includes distributed system, high performance computing,
Blockchain, software development.
\end{IEEEbiography}
\vspace{-30pt}
\begin{IEEEbiography}[{\includegraphics[width=1in,height=1.25in,clip,keepaspectratio]{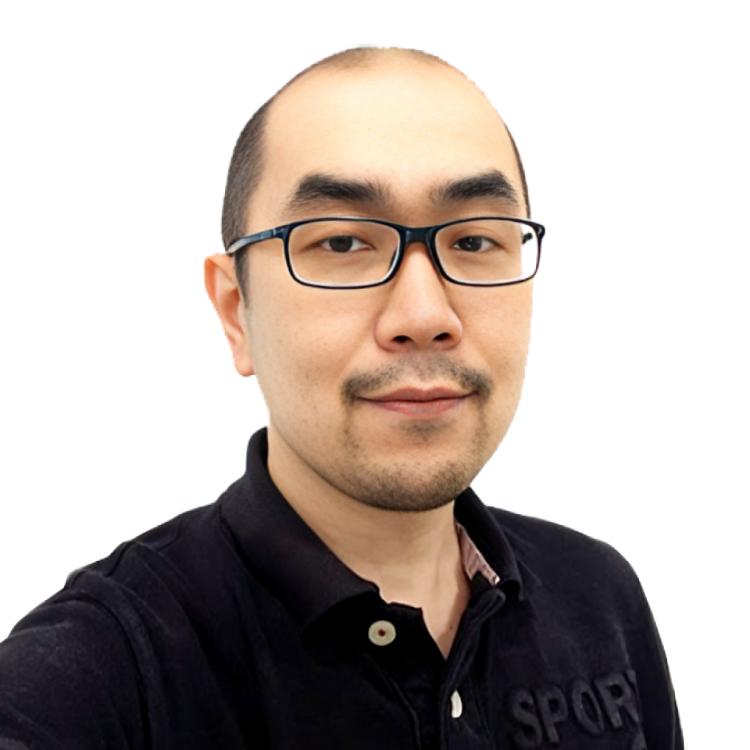}}]{Kerching Choo} is the Co-Founder and Chief Technology Officer of HeLa Labs. He holds degrees in Computer Science from De Montfort University, Business Management from the University of Bradford, and Game Design from Full Sail University. 
He now leads AI and Web3 integration to develop a modular Layer 1 blockchain, advancing innovation in decentralized applications.
\end{IEEEbiography}
\vspace{-30pt}
\begin{IEEEbiography}[{\includegraphics[width=1in,height=1.25in,clip,keepaspectratio]{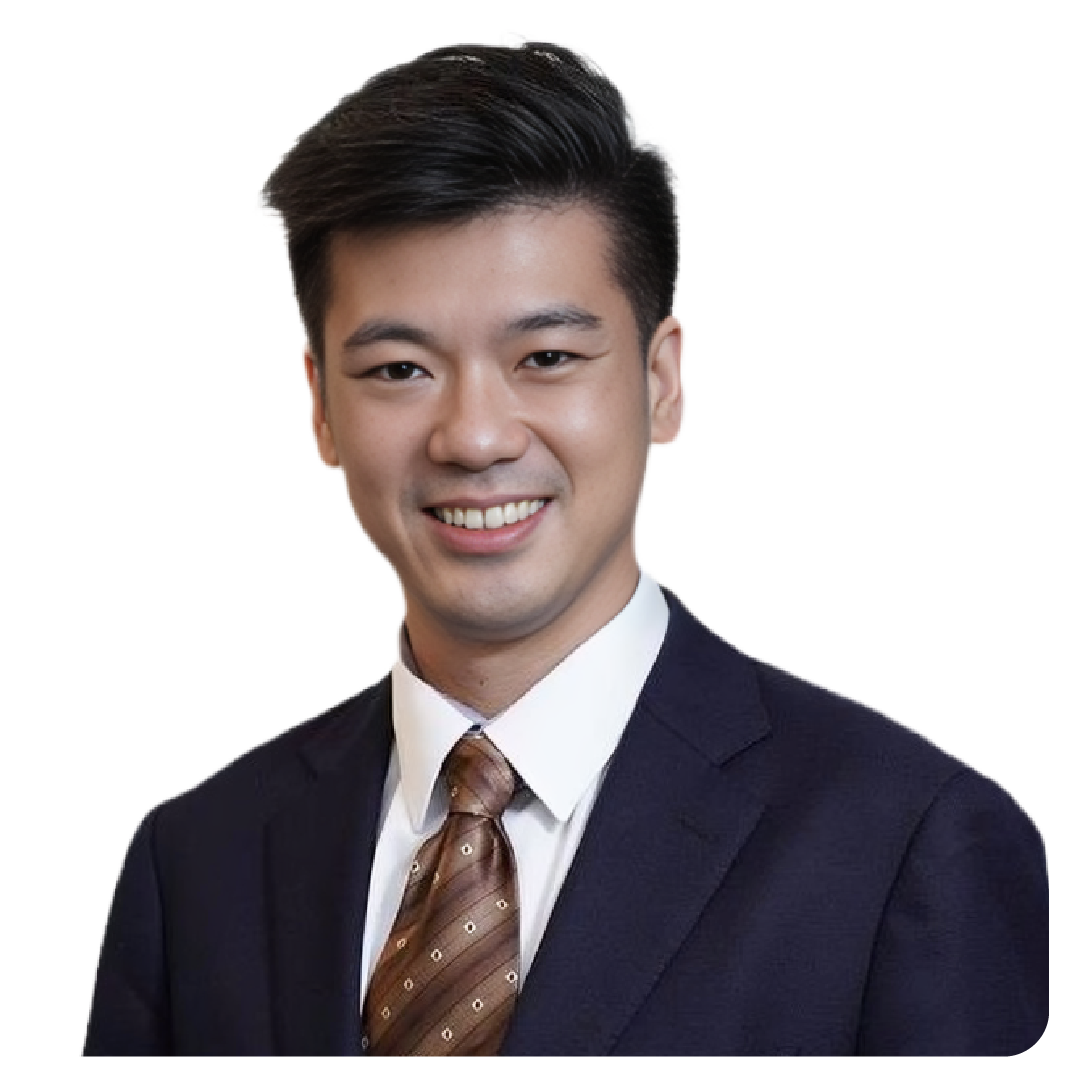}}]{Andy Ting} is the Founder of HeLa Labs, a company focused on building a modular Layer 1 blockchain designed for AI infrastructure across Web2 and Web3.
He holds a PhD in Computer Engineering from Nanyang Technological University.
With a background in high-frequency trading and extensive experience advising banks and financial institutions, Andy’s work focuses on scalable, secure blockchain solutions for industries like gaming, DeFi, and decentralized infrastructure. 
\end{IEEEbiography}
\vspace{-30pt}
\begin{IEEEbiography}[{\includegraphics[width=1in,height=1.25in,clip,keepaspectratio]{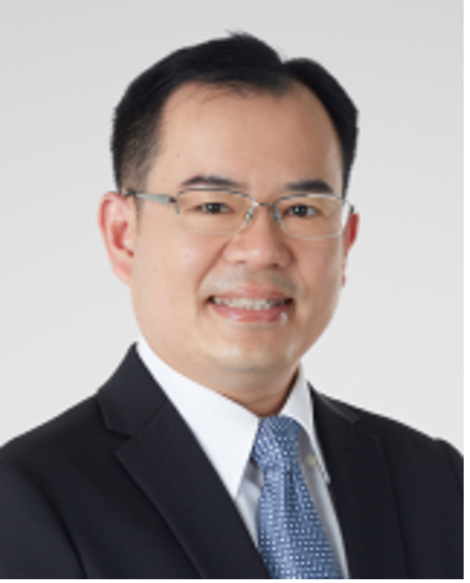}}]{Siow Mong Rick Goh}
received his Ph.D. degree in electrical and computer engineering from the National University of Singapore. He is the Director of the Computing and Intelligence (CI) Department, Institute of High Performance Computing, Agency for Science, Technology and Research, Singapore, where he leads a team of over 80 scientists in performing world-leading scientific research, developing technology to commercialization, and engaging and collaborating with industry. His current research interests include artificial intelligence, high-performance computing, blockchain, and federated learning.
\end{IEEEbiography}
\vspace{-30pt}
\begin{IEEEbiography}[{\includegraphics[width=1in,height=1.25in,clip,keepaspectratio]{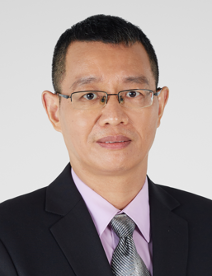}}]{Qingsong Wei}
received the PhD degree in computer science from the University of Electronic Science and Technologies of China, in 2004. He was with Tongji University as an assistant professor from 2004 to 2005. He is a Group Manager and principal scientist at the Institute of High Performance Computing, A*STAR, Singapore. His research interests include decentralized computing, Blockchain and federated learning. He is a senior member of the IEEE.
\end{IEEEbiography}


\end{document}